\documentclass[10pt]{article}
\usepackage{amsfonts}
\usepackage{amssymb}
\usepackage{amsthm}
\usepackage{amsmath}
\usepackage{mathabx} 
\usepackage{color}

\usepackage[pdftex]{graphicx}

\newcommand{\sss}{\setcounter{equation}{0}}
\newtheorem{theorem}{THEOREM}[section]

\newtheorem{lemma}[theorem]{LEMMA}

\newtheorem{corollary}[theorem]{COROLLARY}

\newtheorem{remark}[theorem]{REMARK}

\newtheorem{prop}[theorem]{PROPOSITION}

\newtheorem{definition}[theorem]{DEFINITION}

\newtheorem{assumption}[theorem]{ASSUMPTION}

\def\hr{H_{1,\hbox{\rm rec}}(\Lambda;\ere)}
\def\v{\mathbf v}

\def\hv{\hat{\mathbf v}}

\def\hw{\hat{\mathbf w}}

\def\nb {\mathcal A_{\Phi}(0)}

\newcommand{\dom }{{\rm dom}}

\newcommand{\ere}{ {\mathbb R}}

\newcommand{\ese}{{\mathbb S}}

\def\p2{\mathcal A_{\Phi}(B)}
\def\0p2{\mathcal A_{\Phi}(0)}
\def\sp2{\mathcal A_{\Phi}(B)}
\def\beq{\begin{equation}}
\def\ene{\end{equation}}
\def \ds {\displaystyle}
\newcommand{\bull}{\hfill $\Box$}
\def\qed{\ifhmode\unskip\nobreak\fi\ifmmode\ifinner
\else\hskip5pt\fi\fi\hbox{\hskip5pt\vrule width4pt height6pt
depth1.5pt\hskip1pt}}
\def\v{\mathbf v}

\def\hv{\hat{\mathbf v}}

\def\hw{\hat{\mathbf w}}
\def\curl{\, \hbox{ \rm curl}\,}
\def\mo{\mathbf p}
\def\ta{\tilde{A}}

\def\+out{x^{\rm out}}

\begin{document}

\baselineskip=20 pt
\parskip 6 pt

\title{ {Aharonov-Bohm Effect and High-Momenta Inverse Scattering for the 
Klein-Gordon Equation.}
\thanks{ PACS Classification (2008): 03.65Nk, 03.65.Ca, 03.65.Db, 03.65.Ta.  AMS Classification (2010): 81U40, 35P25
35Q40, 35R30. Research partially supported by the project PAPIIT-DGAPA UNAM IN102215  }}
 \author{ Miguel Ballesteros and Ricardo Weder  \thanks{Fellows of the Sistema Nacional de Investigadores. }   \\
 Departamento de F\'{\i}sica Matem\'atica. \\
 Instituto de Investigaciones en Matem\'aticas Aplicadas y en Sistemas.\\
  Universidad Nacional Aut\'onoma de M\'exico. Apartado Postal 20-126\\ IIMAS-UNAM, Col. San Angel, C.P. 01000, M\'exico D.F., M\'exico\\
miguel.ballesteros@iimas.unam.mx,  weder@unam.mx}

\maketitle

\begin{center}
\begin{minipage}{5.75in}
\centerline{{\bf Abstract}}
\bigskip  
We analyze spin-0 relativistic scattering of charged particles propagating  in the exterior, $\Lambda \subset \mathbb{R}^3$, of a compact obstacle $K \subset \mathbb{R}^3$. The connected components of the obstacle are handlebodies. The particles interact with an electro-magnetic field in $\Lambda$ and an inaccessible magnetic field localized in the interior of the obstacle (through the Aharonov-Bohm effect). We obtain high-momenta estimates, with error bounds, for the scattering
operator that we use to recover physical information: We give a reconstruction method for the electric potential and the exterior magnetic field and prove that, if the electric potential vanishes, circulations of the magnetic potential around handles  (or equivalently, by Stokes' theorem, magnetic fluxes over transverse sections of handles) of the obstacle can be recovered, modulo $2 \pi$.
We additionally give a simple formula for the high-momenta limit of the scattering operator in terms of certain magnetic fluxes, in the absence of electric potential. If the electric potential does not vanish, the magnetic fluxes on the handles above referred can be only recovered modulo $\pi$ and the simple expression of the high-momenta limit of the scattering operator does not hold true.

\end{minipage}
\end{center}

\section{Introduction}\sss

We analyze spin-0 relativistic scattering of charged particles propagating  in the exterior, $\Lambda \subset \mathbb{R}^3$, of a compact obstacle $K \subset \mathbb{R}^3$. The connected components of the obstacle are handlebodies. In particular, they can be the union of a finite number of bodies diffeomorphic to tori or to balls. Some of them can be patched through the boundary. We assume that the particle interacts with a short-range magnetic field $B$ and a short-range electric potential $V$, both of them defined in $\Lambda$. The obstacle is shielded and contains an inaccessible magnetic field. The only information, from the magnetic field inside the obstacle, we may have access to is through circulations of the magnetic potential around the handles of the obstacle. The aim of this paper is proving that the electromagnetic field can be reconstructed from the high-momenta limit of the scattering operator, as well as some information from the circulations of the magnetic potentials around the handles of the obstacle. The latter is the Aharonov-Bohm effect \cite{ab}, \cite{f} a purely quantum phenomenon.      

There are many related results in the literature dealing with a similar setting (obstacle magnetic-scattering and Aharonov-Bohm effect in three dimensions) in the non-relativistic case, see \cite{bw}, \cite{bw2}, \cite{bw3}, \cite{EI2}, \cite{EI1}, and the references quoted there. In this paper we analyze the relativistic case, which is physically and mathematically relevant because all the previous works referred above use a high-velocity limit for their reconstruction formulas. It is, thus, evident the necessity to take into consideration special-relativity if high energies are to be addressed. Regarding the Aharonov-Bohm effect (if the magnetic field vanishes in $\Lambda$), we actually find important differences and similarities between the non-relativistic and the relativistic cases : In the non-relativistic case the leading order of the scattering operator as the velocity $v$  goes to infinity contains only the magnetic potential, and  the contribution of the electric potential appears in the next order term that is of order  $1/v$. However,  in the relativistic case the magnetic and the electric potentials appear in the leading term as the momentum goes to infinity. This means that, in contrast to the non-relativistic case, in the relativistic model the electric and the magnetic potentials have the same order of contribution, which produces some differences in the information one can recover from high-momenta scattering, between both cases, concerning the magnetic potential. Actually, if the electric potential vanishes, we prove that what can be recovered from scattering is pretty much the same in both cases (namely, fluxes of the magnetic potential, modulo $2 \pi$, around the handles) and we find a similar (very simple) expression for the high-momenta limit of the scattering operator, in terms of certain magnetic fluxes. This, however, is not valid anymore if the electric potential does not vanish. In this circumstance, in the relativistic case, the leading term of the high-momenta limit of the scattering operator depends non-trivially on the electric potential and the fluxes around the handles can be recovered only modulo $ \pi$, while in the non-relativistic case having a non-vanishing electric potentials does not change the matter.

 In our work we study inverse-scattering for the Klein-Gordon equation in the case that the sesquilinear form associated with the classical field energy is positive definite. The direct scattering problem in case where it is not positive definite is studied in \cite{Ger}.    

Our main results are presented in Section \ref{mr}. Specifically they are stated in Theorems \ref{TPG}, \ref{inverse-fields}, \ref{th-7.1} and \ref{th-7.12I}. Theorem \ref{TPG} gives the high-momenta limit of the scattering operator. It is the main input from which we recover information from the scattering operator. This is the most laborious result and the core of our proof. The proof of it is presented in Section  \ref{high-long}, which is based in the results of Section \ref{high-short}. As a matter of fact, Sections \ref{main} and \ref{tec} are devoted to the proof of Theorem \ref{TPG}. In Section \ref{main} we give the main arguments, while some technical results are deferred to Section \ref{tec}. In Theorem \ref{inverse-fields} we prove that the electric potential and the magnetic field can be recovered, in a certain region, from the high-momenta limit of the scattering operator. 
Theorem \ref{th-7.1} gives the specific information from the magnetic potential that we can recover from the high-momenta limit of the scattering operator, namely, certain circulations of the magnetic potential around handles of the obstacle. In Theorem \ref{th-7.12I} we provide a very simple expression of the high-momenta limit of the scattering operator in terms of some magnetic fluxes. Theorems \ref{th-7.1} and \ref{th-7.12I} require the electric potential to vanish, since our main interest is to present the Aharonov-Bohm effect. However, similar results are valid in the presence of a non-trivial electric potential, as it is presented in the body of Section \ref{mr} and proved in Section \ref{lala}.

Related results, in two dimensions, are proved in the non-relativistic case in \cite{bw4}, \cite{EI2} and \cite{EI1} (where the long-range behavior of the magnetic potentials is the main issue) and in \cite{n} and \cite{w1}. For relativistic equations in the whole space, see \cite{jung1} and \cite{jung2}. The magnetic Schr\"odinger equation, in the whole space,  is studied in \cite{arians}. The time dependent methods for inverse scattering that we use are introduced in \cite{ew}, for the Schr\"odinger equation. A survey about many different applications of the time dependent method for inverse scattering can be found in \cite{w4}. The direct scattering problem for the Klein-Gordon is studied in \cite{Ger}, \cite{w2}, \cite{w3}, and the references quoted there.

Our paper is organized as follows: Section \ref{dmmr} presents the model and the main results.  In Section \ref{proofs} we give all details of the proofs of our results. It is divided in several subsections: Section \ref{IH} deals with the self-adjointness of the Hamiltonians; Section \ref{wave-scattering} proves the existence of the wave operators and presents some properties of the wave and scattering operators; Section \ref{main} is devoted to the proof of Theorem \ref{TPG}, for the regular case in Section \ref{high-short} and the general case in Section \ref{high-long}. Theorems \ref{inverse-fields}, \ref{th-7.1} and \ref{th-7.12I} are proved, respectively, in Sections \ref{Sinv}, \ref{T2.11} and \ref{T2.12}. In Section \ref{tec} we prove some technical results that are used in Sections \ref{wave-scattering} and \ref{main}. Section \ref{high-stationary} is entirely dedicated to deal with laborious and technical computations used in Section \ref{main}.

\section{Description of the Model and Main Results}  \label{dmmr}
\sss

\subsection{Description of the Model}
\sss We study the propagation of a relativistic  particle outside a bounded  magnet, $K$, in three
dimensions, i.e. the particle  propagates in  the  exterior domain
$\Lambda := \ere^3 \setminus K$. We assume that inside $K$ there is
a magnetic field that produces a magnetic flux. We suppose, furthermore, that in $\Lambda$ there are an
electric potential $A_0$ and a magnetic field $B$. This is a more general situation than the one of the Aharonov-Bohm effect. 
The obstacle $K$ is, of course, a classical macroscopic object defined in $\mathbb{R}^3$. The electromagnetic field is also a classical quantity defined in $\Lambda$, the space where the particles propagate. However, the position of the particle is a quantum quantity which is not represented by the multiplication operator by the variable $x \in \mathbb{R}^3$, where the obstacle lives. As a matter of fact, the components of this operator are not self-adjoint in the Hilbert space at stake and, therefore, they cannot represent a quantum mechanical observable. The free Hamiltonian operator $H_0$ that we describe below (see Definition \ref{freeham})  is diagonalized by a unitary operator $U$ (see \eqref{fw3p}) in such a way that the positive and negative energy subspaces are separated as a direct sum (see \eqref{diagu}). Following \cite{w2}, \cite{w3}, in the  diagonal representation that we just described, we define the position operator as a multiplication operator by the variable $x \in \mathbb{R}^3$. See subsection 2.1.3 for a discussion of these issues.

\subsubsection{The Magnet $K$}
We assume that the magnet $K$ is a compact submanifold of $\ere^3$.
Moreover, $K= {\displaystyle \cup}_{j=1}^L K_j$, where the sets $K_j,  1\leq j \leq L,$ are the
connected components of $K$. We suppose that the $K_j$'s are
handle bodies. For a precise definition of handlebodies see
\cite{bw}, were we study in detail the homology and the cohomology of
$K$ and  $\Lambda$. In intuitive terms, $K$ is the union of a finite
number of bodies diffeomorphic to tori or to balls. Some of them can
be patched through the boundary. See Figure 1.
  
\subsubsection{The Magnetic Field and the Electric Potential}
In the following assumptions we summarize the conditions on the magnetic field and the electric potential that
we use. We denote by $\Delta$ the self-adjoint realization of the Laplacian in $L^2(\ere^3)$ with domain
${\bf  H}^2(\ere^3)$, the Sobolev space of function with distributional derivatives up to order $2$ square integrable. 

\begin{assumption}\label{ass-2.1}
{ \rm We assume that the magnetic field,
 $B$, is a real-valued, bounded $2-$ form in $ \overline{\Lambda}$,
that is two times continuously differentiable  in $\overline  \Lambda$, and, furthermore,
\begin{enumerate}
                                                                                                                                                                                                                                                                                                                                                                                                                                                                                                                                                                                                                                                                                                                                                                                                                                                                                                                                                                                                                                                                                                                                                                                                                                                                                                                                                                                                                                                                                                                                                                                                                                                                                                                                                                                                                                                                                                                                                                                                                                                                                                                                                                                                                                                                        \item
$ B \, \hbox{\rm is closed}:    d B|_{\Lambda} \equiv \mathrm{div}\, B|_{\Lambda} = 0 .$ 

\item
There are no magnetic monopoles in $K$:
\beq
\int_{\partial K_{j}} B=0, \, j \in \{1,2,\cdots,L\}.
\label{2.1}
\ene

\item
\beq
\Big| \Big ( \frac{\partial }{\partial x_1} \Big )^{a} 
\Big (  \frac{\partial }{\partial x_2} \Big )^{b}
\Big (  \frac{\partial }{\partial x_3} \Big )^{c}  B(x)   \Big|  \leq C (1+|x|)^{- \mu}, \, \hbox{\rm for some}\,\, \mu > 2, \: \: \text{and every $a, b, c \in \{0, 1, 2 \} $ 
with $a + b + c \leq 2$},
\label{2.2}
\ene
where $C$ is a general, not specified, constant.

\item
The electric  potential, $A_0$, is a real-valued function defined in $\Lambda$. We suppose that 
 for some $\varepsilon > 0$
\beq \label{pos}
\langle A_0^2\phi  ,  \phi \rangle \leq \langle  \mo \phi,\mo  \phi    \rangle 
+ ( m^2 - \varepsilon ) \langle   \phi,  \phi    \rangle, 
\ene
where $ \mo : = - i \nabla$ is the momentum operator, for every $\phi \in {\bf H}_0^1(\Lambda)$, the closure of $C_0^\infty (\Lambda)$ in ${\bf H}^1(\Lambda)$
(see \cite{chech} and \cite{w2} for explicit conditions on $A_0$ 
implying \eqref{pos}). The latter being the Sobolev space of functions with distributional derivatives up to order $1$ square integrable. In \eqref{pos} we use the inner product in $L^2(\Lambda)$. 
We, furthermore, assume 
there exists a 
constant $C_\delta$ such that  
\beq \label{sub}
\langle A_0^2\phi  ,  \phi \rangle \leq  \delta \langle   \mo  \phi,  \mo  \phi    \rangle 
+ C_\delta \langle   \phi,  \phi    \rangle, 
\ene
for some $\delta < 1/5$ and for every $\phi \in {\bf H}_0^1(\Lambda)$.
 We suppose  additionally that for some $C^{\infty}$
function $\kappa$  defined in $\mathbb R^3$ such that $\kappa(x)=0$ for $ x$ in a neighborhood of $K$, and with $1 - \kappa$ compactly supported, $\kappa A_0$ is two times continuously  differentiable and  
\beq
\Big| \Big ( \frac{\partial }{\partial x_1} \Big )^{a} 
\Big (  \frac{\partial }{\partial x_2} \Big )^{b} \Big (  \frac{\partial }{\partial x_3} \Big )^{c}  \kappa A_0(x)   \Big|  \leq C (1+|x|)^{- \zeta}, \, \hbox{\rm for some}\,\, \zeta > 1, \: \: \text{and every $a, b, c \in \{0, 1, 2 \} $ 
with $a + b + c\leq 2$}.
\label{2.4}
\ene
We notice that the properties of $A_0$ above permit it to have a finite or even an infinite number of singularities.   
\end{enumerate}}
\end{assumption}


\paragraph{The Magnetic Potentials}

Let  $\{ \hat{\gamma}_j\}_{j=1}^m$  be the closed curves defined in  equation   (2.6) of \cite{bw} (see Figure 1). We prove in  Corollary 2.4 of
\cite{bw} that the equivalence classes of these curves
are a basis of the first singular homology group of $\Lambda$. We introduce below a function
that gives the magnetic flux across  surfaces that have  $\{\hat{\gamma}_j\}_{j=1}^m$ as their boundaries.

\begin{definition} \label{def-2.2}{\rm
The flux, $\Phi$, is a function $\Phi: \{\hat{\gamma}_j\}_{j=1}^m \rightarrow \ere$.}
\end{definition}
We now define a class of magnetic potentials with a given flux.

\begin{definition} \label{def-2.3}{\rm
Let $B$ be a closed $2-$ form that satisfies  Assumption \ref{ass-2.1}. We denote by
 $\p2$ the set of all continuous $1-$ forms, $A$, in $ \overline{\Lambda}$ that satisfy
\begin{enumerate}
\item
\beq
|A(x)| \leq C (1+|x|)^{- \zeta}, \qquad \zeta >1,
\label{2.10b}
\ene
\item
\beq
\int_{\hat{\gamma_j}}\, A= \Phi (\hat{\gamma_j}), \hspace{2cm}   \, j \in \{1,2,\cdots, m \},
\label{2.8}
\ene
\item
\beq
d A|_{\Lambda}\equiv \, \curl\, A|_{\Lambda}= B|_{\Lambda}.
\label{2.9}
\ene
\end{enumerate}
Furthermore, we denote by  $\mathcal{A}^{{\rm (reg)}}_\Phi(B) $ the set of $2$-times continuously differentiable functions $A \in \mathcal{A}_\Phi(B)$ such that 
\beq \label{reg}
\Big| \Big ( \frac{\partial }{\partial x_1} \Big )^{a} 
\Big (  \frac{\partial }{\partial x_2} \Big )^{b} \Big (  \frac{\partial }{\partial x_3} \Big )^{c}  A(x)   \Big|  \leq C (1+|x|)^{- \zeta}, \: \: \text{for every $x \in \overline{\Lambda}$ and every $a, b, c \in \{0, 1, 2 \} $ 
with $a + b + c \leq 2$}.
\ene
Here the superscript (reg) stands for regular. 
} 
\end{definition}

\begin{remark} \label{Rreg} { \rm
In Theorem 3.7 of \cite{bw} we construct the Coulomb potential, $A_C$, that belongs to $\sp2$ with  a $\zeta >1$ that depends on $\mu$.
For this purpose condition (\ref{2.1}) is essential. The same proof applies to see that \eqref{2.2} implies that  
$ A_C \in  \mathcal{A}^{{\rm (reg)}}_\Phi(B)  $. Actually, the Coulomb magnetic potential has a regularizing effect, in the sense that it is one time more differentiable that $B$. However, this subtlety is not relevant for the purposes of this paper. Notice that we use the same quantity $\zeta$ in \eqref{2.4}, \eqref{2.10b} and \eqref{reg}. We do it for convenience to keep as simple as possible our notation.   }        
\end{remark}

In  Lemma 3.8 of \cite{bw} we prove that for any $A, \ta \in \p2$ there is a
 $C^1 \, 0-$ form $\lambda$ in $\overline{\Lambda}$
such that
\beq
\tilde{A} - A= d \lambda.
\label{2.11}
\ene
Moreover, we can take 
\beq \label{claro}
\lambda(x):=\int_{C(x_0,x)}(\tilde{A}-A),
\ene
 where $x_0 $
is any fixed
point in $\Lambda$ and $C(x_0,x)$ is any simple
 differentiable curve in $\Lambda$ with starting point   $x_0$ and ending point  $x$. Furthermore, 
\beq \label{lambdainfty} 
 \lambda_\infty(x):=\lim_{r\rightarrow \infty} \lambda(rx) 
\ene 
 exists
 and it is continuous in $  \mathbb{R}^3 \setminus \{ 0 \}  $ and homogeneous of order zero, i.e.
  $\lambda_\infty(r x)=\lambda_\infty(x), r >0,
  x \in  \ere^3 \setminus\{0\}$. Moreover,
\beq
\begin{array}{c}
|\lambda_\infty(x)-\lambda(x)|\leq \int_{|x|}^\infty \, b(|x|), \hbox{\rm for some}\,\, b(r)\in L^1(0,\infty).
\end{array}
\label{2.12}
\ene
Actually, \eqref{2.10b} and \eqref{claro} imply that $ \lambda_\infty $ is a constant function. We denote this constant by $\lambda_\infty \equiv \lambda_{\infty}(x) $.  In Lemma 3.8 of \cite{bw} we consider a more general case where $\lambda_\infty$ is not necessarily constant.

\subsubsection{The Free Klein-Gordon Equation}
 The free Klein-Gordon equation is given by
\beq \label{freekg}
\Big (  i \frac{\partial}{\partial t}  \Big )^2 \phi =  
\Big( {\mathbf p}^2+ m^2 \Big ) \phi, 
\ene
where 
 $m > 0$ is the mass of the particle. 
We do not include the obstacle in the free evolution. This mathematically means that we are looking for solutions $\phi : \mathbb{R}\times  \mathbb{R}^3 \to \mathbb{C}$. To analyze \eqref{freekg} we proceed as in \cite{w2}, \cite{w3} and we study an equivalent system of differential equations that has the advantage of being of order one in $\frac{\partial}{\partial t}$. For this purpose we define: 
\begin{definition}[Free Hamiltonian] \label{freeham}  \rm{
Let $B_0$ be the operator 
\beq \label{B0}
B_0 : = \Big( {\bf p} ^2 + m^2 \Big )^{1/2},  
\ene
with domain ${\bf H}^1(\mathbb{R}^3)$.

We denote by $\mathcal{H}_0$ the Hilbert space 
\beq \label{fhilbertA}
\mathcal{H}_0 : = \dom(B_0) \oplus L^2(\mathbb{R}^3)
\ene
with inner product
\beq \label{finnerA}
\langle \phi, \psi \rangle_{\mathcal{H}_0} : = \langle B_0 \phi_1, B_0 \psi_1  \rangle + 
 \langle  \phi_2, \psi_2 \rangle ,  
\ene
for $\phi = (\phi_1, \phi_2)$, $\psi = (\psi_1, \psi_2)$. Note that under the identification   $ \phi_1=\phi, \phi_2=  \frac{\partial}{\partial t} \phi$,  $ \psi_1=\psi, \psi_2=  \frac{\partial}{\partial t} \psi$,  with $ \phi, \psi$ solutions to \eqref{freekg},   \eqref{finnerA} is the sesquilinear form associated to the classical field energy of the free Klein-Gordon equation. 

The free Hamiltonian $H_0 $ is given by 
\beq \label{ham0}
H_0 : =  \begin{pmatrix}
0  & i \\ -i B_0^2 & 0 
\end{pmatrix},
 \hspace{1cm}  \text{with} \hspace{1cm}  \dom(H(\underline A)) : = \dom(B_0^2) \oplus \dom( B_0) = {\bf H}^2(\mathbb{R}^3) \oplus {\bf H}^1(\mathbb{R}^3).
\ene
Notice that $ H_0  $ is self-adjoint in $\mathcal{H}_0$. 
}
\end{definition}
The free Klein-Gordon equation \eqref{freekg} is equivalent to the system of differential equations
\beq \label{eqham}
i\frac{\partial}{\partial t } \psi = H_0\psi,
\ene 
where $\psi : \mathbb{R} \times \mathbb{R}^3 \mapsto \mathbb{C}^2$ with $ \psi_1=\phi, \psi_2= \frac{\partial}{\partial t} \phi$. Note the slight difference with \cite{w2}, \cite{w3} where the reduction to a  system is made with $ \psi_1= \phi, \psi_2= i \frac{\partial}{\partial t} \phi$.

Let us denote,
$$
L^2\left( \mathbb R^3 \right)^2:= L^2\left( \mathbb R^3\right)\oplus  L^2\left( \mathbb R^3\right),
$$
 and consider the following unitary transformation :  
\beq \label{fw}
F_W : \mathcal{H}_0 \mapsto   L^2(\mathbb{R}^3)^2 
\ene
given by 
\beq \label{fwm}
F_W : = \begin{pmatrix}
B_0 & 0 \\ 0 & 1
\end{pmatrix}.
\ene
It follows that 
\beq\label{fw1}
F_W H_0 F_W^{-1} = B_0 \beta,
\ene
where 
\begin{align}\label{a5}
\beta : = \begin{pmatrix}  0  & i \\ -i & 0  \end{pmatrix}.
\end{align}
We define the matrices 
\beq \label{fw2}
 Q : = 2^{-1/2}\begin{pmatrix}
1 & i \\ 1  & -i
\end{pmatrix}, \hspace{1cm} Q^{-1} := 2^{-1/2}\begin{pmatrix}
1 & 1 \\ -i  & i
\end{pmatrix}, 
\ene
that diagonalize $ \beta $ :
\beq \label{fw3}
Q \beta  Q^{-1} = \begin{pmatrix}
1 & 0 \\ 0 & - 1.
\end{pmatrix},
\ene
Let $U$, \cite{w2}, \cite{w3} be the unitary operator  
\beq \label{fw3p}
U : = Q F_W, \qquad \mathcal {H}_0 \mapsto   L^2(\mathbb{R}^3)^2 .
  \ene
It follows that
\beq \label{diagu}
\hat{H}_0:=U  H_0U^{-1} = \begin{pmatrix}
  B_0  & 0 \\ 0 & -   B_0 
\end{pmatrix}.
\ene
In this representation the free Klein-Gordon equation is equivalent to the system,
\beq \label{rep}
i\frac{\partial}{\partial t}\, \psi= \hat{H}_0\, \psi, \qquad \psi= \left(\begin{array}{l} \psi_+\\ \psi_-
\end{array}\right) \in  L^2(\mathbb{R}^3)^2 .
\ene
The functions $\psi_+, \psi_-$ are, respectively, the positive and negative energy components of the solution. The negative energy solutions are interpreted as antiparticles, in the usual way. In the physics literature, the representation 
\eqref{rep}, but with a scalar product that is not positive definite, is called the free particle or Feshbach-Villars representation, see \cite{fv}, \cite{gr}. We define the position operator \cite{w1}, $ \hat{x}_P$ as multiplication by $x$ in  this representation,

\beq\label{pos1}
\hat{x}_P\, \psi(x)= x\, \psi(x),
\ene           
and then, $\left| \psi_{\pm} \right|^2$ are, respectively, the probability densities for particles with positive and negative energy. Note that this is possible because the scalar product in $ L^2(\mathbb{R}^3)^2 $ is positive definite.

In the representation in $\mathcal{H}_0$ the position operator is given by,
\beq\label{pos2}
x_P:= U^{-1}\, \hat{x}_P\, U= \begin{pmatrix}\displaystyle
  x+ i \frac{\mathbf{p}}{B_0^2}  & 0 \\ 0 & x 
\end{pmatrix}.
\ene
Note that  $x+ i \frac{\mathbf{p}}{B_0^2}$ is different from the Newton-Wigner position operator \cite{nw}.

Observe that multiplication by $x$ in the representation $\mathcal{H}_0$ can not be a position operator. In fact, it is not a selfadjoint operator in $\mathcal{H}_0$ and, hence, it is not a quantum mechanical observable. Actually in the representation $\mathcal{H}_0, x$ is a classical parameter that is used to parametrize the classical, macroscopic, objects like the magnet $K$ and the electric and magnetic fields, but, as mentioned above, the operator that gives the position of the quantum particle is $x_P$.   
 
\noindent {\bf High-Momentum States}

We designate by $ \mathbb{S}^2$ the unit sphere in $ \mathbb R^3$. We need consider high-momentum states that under the free evolution have negligible interaction with the magnet. 
For this purpose,  for every $ \nu \in \mathbb{S}^2$ we denote by (see Eq. \eqref{2.4})  
\beq \label{Lambdanu1}
\Lambda_{ \nu} : = \Big \{ x \in \Lambda \, : \, x + \tau \nu \in \Lambda, \:  \forall \tau \in \mathbb{R} \Big \}, 
\ene
 and

\beq \label{Lambdanu}
\Lambda_{\kappa, \nu} : =
\Big \{x \in \Lambda_{ \nu}  \, : \, \kappa(x + \tau \nu) = 1, \:  \forall \tau \in \mathbb{R}  \Big   \}. 
\ene
  Since the classical free evolution of a relativistic particle is given by $ x + \tau \nu$ for some $\nu \in  \mathbb{S}^1$,
  a state that in the representation $ L^2\left(\mathbb R^3  \right)^2$ is given by a function  $ \phi$ with support in $\Lambda_\nu$
  has no interaction with the magnet under the classical evolution since,
  
  $$
  \left(\chi_{K}(x_P)\, U^{-1} \phi(\cdot+ \nu \tau), U^{-1}\, \phi(\cdot+ \nu \tau)\right)_{\mathcal H_0}=
   \left(\chi_{K}(x)\, \phi(x+ \nu \tau),\, \phi(x+ \nu \tau)\right)_{L^2\left(\mathbb R^3  \right)^2}=0,
  $$ 
where for any set $O$ we denote by $\chi_O$ the characteristic function of $O$.
Our high-momenta states are defined in the representation   $ L^2(\mathbb{R}^3)^2 $   as,
\beq\label{highmom1}
 e^{ix\cdot v\nu}\, \phi(x), \qquad \phi \in L^2\left(\mathbb R^3  \right)^2, \, \mathrm{with} \, \phi \,  \mathrm{supported \, in}\, \Lambda_\nu.
 \ene 
In the representation  $\mathcal H_0$ they are given by,
$$
e^{ix_P\cdot v\nu}\, U^{-1}\,\phi(x), \qquad \phi \in L^2\left(\mathbb R^3  \right)^2, \, \mathrm{with} \, \phi  \,\mathrm{supported \, in}\,  \Lambda_\nu.
$$

 In Eq. \eqref{highmom1} the operators $ e^{ i x \cdot v \nu } $ represent a momentum shift corresponding to $v \nu$. Then $\ v \nu$ symbolize momentum. We use the notation $v$ to represent the norm of the shifted momentum. We proceed in this way to keep a   notation similar to the one  we used in previous papers (\cite{bw}-\cite{bw4}), where the non-relativistic case is addressed. The high-momentum limit amounts to take  $v$ to infinity. The physical intuition is that for high momentum the free quantum evolution is close to the classical free evolution and then, our high momenta states will have negligible interaction with the magnet $K$.

\subsubsection{The Interacting Klein-Gordon Equation}

 The Klein-Gordon equation for a particle in $\Lambda$  with electric potential $ A_0$ and magnetic field $ B$   is given by
\beq
\Big (  i \frac{\partial}{\partial t} -  A_0 \Big )^2 \phi =  
\Big( ({\mathbf p}-  A)^2+ m^2 \Big ) \phi,
\label{2.13}
\ene
where the magnetic potential $A \in \mathcal A _{\Phi}(B)$
satisfies $ d   A=  B$ and $\phi : \mathbb{R}\times \Lambda \mapsto \mathbb{C}$ is the wave function.  As in the free case,
to analyze (\ref{2.13}), we trade it by an equivalent system of differential equations of order $1$ in the time derivative \cite{w1}, \cite{w2}: 
\beq \label{eqham1}
i\frac{\partial}{\partial t } \psi = H(\underline A)\psi,
\ene
where $\psi : \mathbb{R} \times \Lambda \mapsto \mathbb{C}^2$, 
\beq \label{ua}
\underline A : = (A_0, A)
\ene
and 
\beq \label{ham}
H(\underline{A}) : =  \begin{pmatrix}
0  & i \\ -i B(\underline A)^2 & 2 A_0
\end{pmatrix},
 \hspace{2cm}  \text{with} \: \: B(\underline A)^2 
: = \Big( {\bf p} - A \Big )^2 + m^2  - A_0^2. 
\ene
Equation \eqref{2.13} is equivalent to \eqref{eqham1}  with $ \psi_1=\phi, \psi_2= \frac{\partial}{\partial t} \phi$.

In Section \ref{IH} we prove that $ B(\underline A)^2 $ is a strictly
positive operator and that the interacting Hamiltonian, $ H(\underline{A}) $, whose domain is described below, is a self-adjoint operator.     

\begin{definition}[Interacting Hamiltonian] \label{Dham} \rm{
We denote by $\mathcal{H}(\underline A)$ the Hilbert space 
\beq \label{hilbertA}
\mathcal{H}(\underline A) : = \dom(B(\underline A)) \oplus L^2(\Lambda)
\ene
with inner product
\beq \label{innerA}
\langle \phi, \psi \rangle_{\mathcal{H}(\underline A)} : = \langle B(\underline A)\phi_1, B(\underline A)\psi_1  \rangle + 
 \langle  \phi_2, \psi_2 \rangle ,  
\ene
for $\phi = (\phi_1, \phi_2)$, $\psi = (\psi_1, \psi_2)$. The domain of the operator 
$H(\underline A)$ is given by 
\beq \label{domHA}
\dom(H(\underline A)) : = \dom(B(\underline A)^2) \oplus \dom( B(\underline A)). 
\ene }
\end{definition} 
\noindent 
Notice that the specific properties of the electromagnetic potential we choose imply that
$$
 \langle \phi, \psi \rangle_{\mathcal{H}(\underline A)} \geq
\varepsilon \, \langle \phi, \psi \rangle_{L^2(\mathbb{R}^3)^2}, 
$$ 
for every $\phi, \psi \in \mathcal{H}(\underline{A})$ (see  Section \ref{min} below).  The scalar product \eqref{innerA} is the sesquilinear form associated with the classical field energy of the interacting Klein-Gordon equation \eqref{2.13}. In \cite{w2}, \cite{w3} (see also Subsection \ref{IH}) it is proven that \eqref{2.13}  can be represented in $ L^2(\mathbb{R}^3)^2 $ as a first order in time system,  as in the free case.
\subsubsection{Wave and Scattering Operators}

Let $J$ be the identification operator from $\mathcal H_0$ onto $\mathcal H(\underline A)$ given by 
\beq\label{J}
J : = \begin{pmatrix} 
B(\underline A)^{-1} \chi_\Lambda B_0 & 0 \\ 0 & \chi_\Lambda 
\end{pmatrix},
\ene
here $\chi_{(\cdot )}$ is the characteristic function of  $(\cdot)$.
The wave operators are defined as follows:
\beq
  W_{\pm}(\underline A):= \hbox{\rm s-}\lim_{t \rightarrow \pm \infty} e^{it
H(\underline A)}\, J\, e^{-it H_0},
\label{2.19}
\ene
provided that the limits exist. In Section \ref{exwave} we prove the existence of the limits, for every  $\underline A = (A_0,A)$,  with $ A \in \mathcal{A}_{\Phi}(B)$. 

The scattering operator is defined, for every $\underline A = (A_0,A)$  with $ A \in \mathcal{A}_{\Phi}(B)$, by 
\beq
S(\underline A) = W_+^*(\underline A) W_-(\underline A). 
\ene

\subsection{Main Results} \label{mr}

\subsubsection{Notation}

 \begin{definition}\label{WUA} \rm{
Let $ \underline A = (A_0, A)$,  with  $ A \in A_{\Phi}(B)$.  
For every $\nu \in \mathbb{S}^2$
we define (see \eqref{2.4})
\begin{align}\label{i33}
W_{\underline A}(t,  \nu , \kappa ) = e^{i t \beta \nu \cdot \mo  }\begin{pmatrix} \kappa A_{0}(x) &  -i  \kappa A \cdot \nu   \\
i  \kappa A \cdot \nu  & \kappa A_{0}(x) \end{pmatrix} e^{-i t \beta \nu \cdot \mo } .
\end{align} }
\end{definition}
\noindent Notice that 
\begin{align} \label{in1t}
   e^{-i \int_{-\infty}^{\infty} W_{\underline A}(r, v\nu, \kappa) dr }  
  = 
   Q^{-1} \begin{pmatrix}
 e^{i \int_{-\infty}^{\infty} dr  (   A^{(j)}\cdot \nu - A_0) (x + r \nu ) }  & 0 
\\ 0 & e^{- i \int_{-\infty}^{\infty} dr   ( A^{(j)}\cdot \nu + A_0) (x + r \nu ) }
\end{pmatrix} Q  ,  
\end{align}
which follows from \eqref{cas}-\eqref{in2}. \\

\subsubsection{High-Momenta Limit of the Scattering Operator}
The theorem below gives the high-momenta limit of the scattering operator, from which we reconstruct the electric potential, the magnetic field and some properties of the magnetic potentials. The derivation of this formula is the most laborious part in our paper and the core of our proofs. The proof of this theorem is deferred to Section \ref{high-long}, which is based in the results of Section \ref{high-short}.  For the definition of the weighted Sobolev space   $ {\bf H}_{\langle x \rangle^{4 l}}^2(\mathbb{R}^3)^2$ see Subsection 3.1.

\begin{theorem} \label{TPG}
Set $\nu \in  \mathbb{S}^2 $ and $l \in \mathbb{N}$,  $l \geq \zeta/2, l \geq 2$.
Suppose that $ \phi, \psi \in  {\bf H}_{\langle x \rangle^{4 l}}^2(\mathbb{R}^3)^2$
are supported in $\Lambda_{\kappa, \nu} $. Let $ \underline A = (A_0, A)$,  with  $ A \in A_{\Phi}(B)$.
Then  
\begin{align} \label{TP1I}
\langle e^{-i x \cdot v \nu } U  S (\underline{A})  U^{-1} e^{i x \cdot v\nu }\phi \; , 
\psi  \rangle_{L^2(\mathbb{R}^3)^2} =  &   \Big \langle  \begin{pmatrix}
 e^{i \int_{-\infty}^{\infty} dr  (   A^{(j)}\cdot \nu - A_0) (x + r \nu ) }  & 0 
\\ 0 & e^{- i \int_{-\infty}^{\infty} dr   ( A^{(j)}\cdot \nu + A_0) (x + r \nu ) }
\end{pmatrix}  \phi,\; \psi \Big \rangle  ,
 \\ & +  \|    \phi   \|_{ {\bf H}_{\langle x \rangle^{4 l}}^2 (\mathbb{R}^3)}  
 \|    \psi   \|_{{\bf H}_{\langle x \rangle^{4 l}}^2 (\mathbb{R}^3)}
 \begin{cases} O\Big ( v^{1 - \zeta}  + \frac{1}{v} \Big ) & \text{if} 
 \: \zeta \ne 2 \\ \\
 O\Big ( \frac{ \ln(v)}{v}  \Big ) & \text{if} 
\: \zeta = 2. 
 \end{cases} \notag
\end{align}

\end{theorem}
In Remark \ref{rempw} we additionally derive high-momenta expressions for the wave operators. Notice that Theorem \ref{TPG} does not require the functions $\phi, \psi$ to be compactly supported, as it is done in our previous works (see \cite{bw}-\cite{bw4}). The function $\kappa$ is introduced for two reasons: To cutoff smoothly the obstacle and to cutoff the singularities of $A_0$. If $A_0$ has no singularities, then our result is equivalent to the results in \cite{bw}-\cite{bw4}, for the non-relativistic case: If $A_0$ has no singularities, and $ \phi, \psi $ are supported in $\Lambda_\nu$, we can always find   some $C^{\infty}$ function $\kappa$  defined in $\mathbb R^3$ such that $\kappa(x)=0$ for $ x$ in a neighborhood of $K$, and with $1 - \kappa$ compactly supported,  such that $ \phi, \psi $ are supported in $\Lambda_{\nu, \kappa}$.

Similar results for non-obstacle relativistic-scattering are presented in \cite{jung1}, \cite{jung2}. There error bounds are not provided and a different class of (bounded) electromagnetic potentials is addressed.             

\subsubsection{Inverse-Scattering Reconstruction Method}
In this section we present one of our main results, namely  Theorem \ref{inverse-fields}. The proof is postponed to Section \ref{Sinv}.

\begin{definition} \label{lrec}{\rm
We denote by $\Lambda_{{\rm Rec}}$ the set of points $ x \in \Lambda $ such that, for some two-dimensional plane $P_x$, $x + P_x \subset  \left(\kappa^{-1}(\{ 1\})\right)^\circ $, for  some $C^{\infty}$
function $\kappa$  defined in $\mathbb R^3$ such that $\kappa(x)=0$ for $ x$ in a neighborhood of $K$, and with $1 - \kappa$ compactly supported, where for any set $O$ we denote by $O^\circ$ its interior.}
 
\end{definition}

\begin{theorem} \label{inverse-fields}
The high-momenta limit \eqref{TP1I} of the scattering operator uniquely determines $B(y)$ and $A_0(y)$ for every $y \in \Lambda_{Rec}$. 
\end{theorem}
\begin{remark}\label{TRW} {\rm
Notice that in Definition \ref{lrec} the functions $\kappa$ are not fixed, but can be conveniently selected.  Furthermore, in the proof of Theorem  \ref{inverse-fields} in Subsection 3.5.1 we give a method for the  unique reconstruction of $ B(y), A_0(y), y \in |\Lambda_{\mathrm{Rec}}$.   }
\end{remark}

\subsubsection{The Aharonov-Bohm Effect}
In this section we assume that $B =  0$ and $A_0 = 0$, i.e., that the electromagnetic field vanishes in $\Lambda$. The hypothesis $A_0 = 0$ is assumed for convenience, in the spirit of the Aharonov-Bohm effect. Nevertheless some results are also valid for $A_0 \ne 0$. In the case that $A_0 \ne 0$, notable differences and similarities  between the relativistic and the non-relativistic cases hold true. They are presented in Section \ref{lala}, where the corresponding results for $A_0 \ne 0$  are proved.
 
\paragraph{Notation}

For any $x \in \Lambda_{\hv}$ and any unit vector $\hv \in \ese^2$ we denote
\begin{equation} \label{line}
L(x,\hv):= x+\ere \hv,
\end{equation}
and we give to $L(x,\hv)$ the orientation of $\hv$. Suppose that $x\in \Lambda_{\hv}, y \in \Lambda_{\hw}$, $\hv, \hw \in \ese^2$
satisfy $ \hv\cdot\hw \geq 0$ and that
$$
L(x,\hv) \cup L(y,\hw) \subset \Lambda.
$$
Take $ \rho >0$ so large that
$$
\hbox{\rm convex}\, \left( (x+ (-\infty, -\rho] \hv ) \cup  (y+ (-\infty, -\rho] \hw )\right)
\cup \,\hbox{\rm convex}\, \left( (x+ [\rho, \infty) \hv ) \cup  (y+ [\rho, \infty, ) \hw )\right)
\subset \mathbb{R}^3 \setminus {B(0; r)},
$$
where $ K \subset B(0; r)$ and
 the symbol convex$(\cdot)$ denotes the convex hull of the indicated set.

We denote by $\gamma(x,y,\hv,\hw)$ the continuous, simple, oriented  and closed curve
with sides $ x+[-\rho, \rho]\hv$, oriented in the direction of $\hv$,
 $ y+[-\rho, \rho]\hw$, oriented in the direction of $-\hw$, and the oriented straight lines that join the
 points $x+\rho \hv$ with $y+ \rho\hw$ and $y-\rho \hw$ and $x-\rho \hv$.

\paragraph{Results}

\begin{theorem} \label{th-7.1}
Suppose that $B = 0$ and that $A_0 = 0$. Then, for any flux, $\Phi$,
and all $ A\in \mathcal{A}_\Phi(0)$, the high-momenta limit of  $S(\underline A$)  in  \eqref{TP1I}, known  for $\hv$ and $\hw$, 
determines the fluxes
\beq
 \int_{\gamma (x,y,\hv,\hw)}A
\label{7.3}
\ene
modulo $2\pi$, for all curves $\gamma (x,y,\hv,\hw)$.
\end{theorem}
Theorem \ref{th-7.1} is also valid if $A_0 \ne 0$, with the restriction that \eqref{7.3} is determined only modulo $ \pi$ (see Section \ref{lala}). The proof of Theorem \ref{th-7.1} is done at the beginning of Section \ref{San}. This Theorem gives additionally information about the de Rham cohomology class of $A$, see Remarks \ref{rem-7.2} and \ref{rm-7.3}.   

\begin{theorem}\label{th-7.12I} Suppose that $A_0 = 0$.  There is an open disjoint cover of $\Lambda_\nu$, 
$\big \{ \Lambda_h  \big \}_{h \in \mathcal{I}}  \bigcup \{ \Lambda_{{\rm out}}  \} $, and a set of real numbers $ \big \{ F_h  \big \}_{h \in \mathcal{I}} $ 
such that the following holds true:
 Set  $\phi, \psi \in  {\bf H}_{\langle x \rangle^{4 l}}^2(\mathbb{R}^3)^2$ as in Theorem \ref{TPG}, with $\phi$ compactly supported. 
For every $A \in \nb$   
\begin{align}\label{7.10In}
\langle U S(\underline A) U^{-1}\,  e^{i v \nu \cdot x}\phi, e^{i v\nu \cdot x }\psi \rangle_{L^2(\Lambda)^2} = &  \Big \langle   \,\left( \sum_{h\in \mathcal{I}}\,
 \begin{pmatrix}  e^{i F_h}  & 0 \\ 0 &   e^{-i F_h} \end{pmatrix}  \chi_{\Lambda_h} \phi + 
  \chi_{\Lambda_{{\rm out}} }\phi \right)\: ,  \psi \Big \rangle_{L^2(\Lambda)^2}\\ \notag & + O\left( \frac{1}{v}\right) \|    \phi   \|_{{\bf H}_{\langle x \rangle^{4 l}}^2 (\mathbb{R}^3)}  
 \|    \psi   \|_{{\bf H}_{\langle x \rangle^{4 l}}^2 (\mathbb{R}^3)} ,
\end{align}
as $v $ tends to infinity. 
\end{theorem}
The sets $ \big \{ \Lambda_h  \big \}_{h \in \mathcal{I}} $ have a geometric meaning, they are the holes of $K$ in the direction $\nu$. The numbers  $ \big \{ F_h  \big \}_{h \in I} $  are magnetic fluxes (around handles of the obstacle) associated to the holes $h$, for $h \in \mathcal{I}$.  
 The set $\Lambda_{{\rm out}}$ is the region without holes. This is well explained in the lines above Theorem \ref{th-7.12}. The proof of 
 Theorem \ref{th-7.12I} is done at the end of Section \ref{San}, this theorem is actually rephrased in Theorem \ref{th-7.12}. The simple expression (\ref{7.10In}) is not anymore valid if $A_0 \ne 0$ (see Section \ref{lala}), whereas in this case it is valid in the non-relativistic setting (see \cite{bw}).    
\begin{remark} {\rm
Suppose that $A_0 = 0$ and  that  $\phi$ and $\psi$ are compactly supported in $\Lambda_{\nu }$, then we can always find
some $C^{\infty}$ function $\kappa$  defined in $\mathbb R^3$ such that $\kappa(x)=0$ for $ x$ in a neighborhood of $K$, and with $1 - \kappa$ compactly supported, such that $\phi$ and $\psi$ are compactly supported in $\Lambda_{\kappa, \nu} $. It follows that if $\phi$ and $\psi$ are compactly supported in $\Lambda_{\nu }$ and belong to $   {\bf H}^2(\mathbb{R}^3)^2 $,
the conclusions of Theorem \ref{th-7.12I} are valid.   }
\end{remark}

\section{The Proofs}\label{proofs}
\sss
\subsection{General Notation}
We denote by $C$ any finite positive constant whose value is not
specified. For any $ x\in \ere^3, x \neq 0$, we denote, $\hat{x}:=
x/|x|$.  For every $ \v \in \ere^3$ we designate $ v:= |\v|$. 
The domain of an operator or a quadratic form is denoted by 
$\dom(X)$,
where either $X$ is an operator or a quadratic form.  For every open set 
$O\subset \mathbb{R}^3$, we use the symbol 
$\langle \cdot , \cdot \rangle $
to represent the inner product in $L^2(O)$. If it is clear from the context, we use the same symbol to represent the inner product in  $L^2(O)^2: = L^2(0)\oplus L^2(O)$ (in general, for every Hilbert space $ \mathcal{H} $, we set $\mathcal{H}^2 : = \mathcal{H} \oplus \mathcal{H}$).  For a general (complex) Hilbert space $\mathcal H$, its inner produc is represented by
$ \langle \cdot, \cdot  \rangle_{\mathcal{H}} $.  
The norm of an element $x $ in a normed vector space $X$ is denoted by 
$  \|\cdot \|_X  $. However, if no confusion arises, we omit in general the subscript in the case that $ X $ is $L^2 (O)$, $L^2(O)^2$ or a space of operators with the corresponding operator norm.  \\
In this paper we use a system of units in which the numerical value of the 
charge of the particle, the speed of light and the Plank's constant are one:   
$e = 1,  c = 1,  \hbar = 1 $. 
The Schwartz space of rapidly decreasing $C^\infty $-functions in $\mathbb{R}^3$ is denoted by 
$\mathcal{S}(\mathbb{R}^3)$.
For every $n \in \mathbb{N}$ and every open set $ O $ in $\mathbb{R}^3$, we denote by 
${\bf H}^n(O)$ 
the Sobolev  space of functions with distributional derivative up to order $n$ square integrable, and by 
${\bf H}_0^n(O)$
the closure of $C_0^\infty(O)$ in ${\bf H}^n(O)$, see \cite{adams}. For every strictly  positive function $\omega : O \mapsto \mathbb{R} $ we denote by $ {\bf H}^n_{\omega}(O) $ the corresponding weighted Sobolev spaces. For every 
$\phi \in {\bf H}^n_{\omega}(O) $, 
$$
\| \phi \|_{{\bf H}^n_{\omega}(O)} := \sum_{\alpha_1 + \alpha_2 + \alpha_2 \leq n} \Big \|  \omega^{1/2} \frac{\partial^{\alpha_1} }{\partial x_1^{\alpha_1} }\frac{\partial^{\alpha_2} }{\partial x_2^{\alpha_2} }\frac{\partial^{\alpha_3} }{\partial x_3^{\alpha_3} } \phi  \Big \|_{L^2(O)}. 
$$
  
%
%
%
%
%
%
\noindent The open ball in $\mathbb{R}^3$ of radius $r$ and center $x$
is denoted by
$B(x; r)$,
and for every set $O$, its complement is denoted by $O^{\displaystyle{c}}$. 
The Fourier transform of a function $\phi \in L^2(\mathbb{R}^3)$ is denoted by  
\beq \label{fou}
\widehat{\phi}(p) \equiv \mathcal{F}(\phi)(p) : = \frac{1}{(2\pi)^{3/2}} \int
e^{- i x \cdot p } \phi(x)dx 
\ene
and the inverse Fourier transform by
\beq \label{ifou}
\widecheck{\phi}(x) \equiv \mathcal{F}^{-1}(\phi)(x)   : = \frac{1}{(2\pi)^{3/2}} \int
e^{ i x \cdot p } \phi(p)dp .
\ene  
Another useful notations that we adopt are the symbols 
$ \langle x \rangle := (1 + x^2)^{1/2}, \langle \mo \rangle := (1 + \mo^2)^{1/2}$,
for $x \in \mathbb{R}^3$ and $\mo = -i \nabla$. In this paper we use
use the standard identifications between differential forms in open subsets of $\mathbb{R}^3$ and vector calculus (see page 348 in \cite{bw}, for example). \\
A useful formula that we use very much in this paper is the following: For every measurable function 
$f: \mathbb{R}^3 \mapsto \mathbb{C}$
\beq \label{usefullEq}
e^{-i x \cdot v \nu } f(\mo ) e^{i x \cdot v \nu} = f (\mo + v\nu),
\ene
for every $v \geq  0$ and $\nu \in \mathbb{S}^2$.   
 
\subsection{The Interacting Hamiltonian}\label{IH}

In this section we prove that $H(\underline A)$ is a selfadjoint operator and give a proper definition of $B(\underline A)$. Related results are presented in \cite{w2} and \cite{w3}.

Let $   q_{\underline A} : C_0^\infty (\Lambda) \times C_0^\infty (\Lambda)\to \mathbb{C} $ be the bilinear form defined by
\beq
q_{\underline A}(\phi,\psi):= \Big \langle (\mo - A) \phi,(\mo - A)\psi \Big \rangle + m^2 \langle \phi, \psi \rangle 
- \langle A_0^2 \phi, \psi \rangle.  
\label{2.15}
\ene
The diamagnetic inequality (see Lemma 1.2, Chapter 9 in \cite{chech}) and \eqref{pos} imply that 
$\label{below}
q_{\underline A} \geq \varepsilon. 
$
Theorem X.17 in \cite{rs2} (KLMN theorem) implies that $q_{\underline A}$ is closable. We denote its closure by 
$
\overline q_{\underline A},
$
it is represented by a selfadjoint  operator
$\label{min}
B(\underline A)^2 \geq \varepsilon. 
$ 
Set $ B(\underline A)$ the positive square root of $B(\underline A)^2$. It follows that 
\beq \label{domq}
\dom(B(\underline A)) = \dom(\overline q_{\underline A}), \hspace{2cm} \overline q_{\underline A}(\phi, \psi) = 
\big ( B(\underline A) \phi, \: B(\underline A)\psi \big ),
\ene
for every $\phi, \psi \in \dom(\overline q_{\underline A})$. Actually, \eqref{sub} and the fact that  is $A$ is bounded imply that
\beq \label{domq1}
\dom(\overline q_{\underline{A}}) = {\bf H}_{0}^1(\Lambda). 
\ene
\begin{prop}
 $ H(\underline A)  $ is a self-adjoint operator (see Definition \ref{Dham}).  
\end{prop}
\noindent{\it Proof:}
Defining the unitary operator $ U_{\underline A}: \mathcal{H}(\underline A) \mapsto  L^2(\Lambda)^2 $
by 
\beq \label{pr1}
\psi \mapsto Q \begin{pmatrix}
B(\underline A) & 0 \\ 0 & 1
\end{pmatrix} \psi
\ene
we find that 
\beq \label{pr2}
  \begin{pmatrix}
0  & i \\ -i B(\underline A)^2 & 0
\end{pmatrix} = U_{\underline A}^{-1}  \begin{pmatrix}
 B(\underline A)& 0 \\ 0 & -  B(\underline A)
\end{pmatrix} U_{\underline A},
\ene
which implies that 
$$    \begin{pmatrix}
0  & i \\ -i B(\underline A)^2 & 0
\end{pmatrix} 
$$
 is self-adjoint, with domain $\dom(H(\underline A))$. 
 
 By \eqref{sub},
 $$
 \left\|  A_0 \phi \right\|^2 \leq \delta  \left\|  {\mathbf p} \phi \right\|^2 + C_\delta  \left\| \phi \right\|^2 \leq \delta\,\overline{q}_{\underline{A}}(\phi,\phi)+ \delta    \left\|  A_0 \phi \right\|^2 +C_\delta  \left\| \phi \right\|^2.
 $$
 Then,
 $$
  \left\|  A_0 \phi \right\|^2 \leq \tilde{\delta} \, \overline{q}_{\underline{A}}(\phi,\phi) + C_{\tilde{\delta}}  \left\| \phi \right\|^2,
  \quad  \tilde{\delta} := \frac{\delta}{1-\delta},\,  C_{\tilde{\delta}} := \frac{C_\delta}{1-\delta},
  $$
 where we use the diamagnetic inequality (Lemma 1.2, Chapter 9 in \cite{chech}).
 Let $\psi = (\psi_1, \psi_2) \in  \dom(H(\underline A))$, then,


\begin{align} \label{pr5}
 \Big \|    \begin{pmatrix}
0 & 0 \\ 0 & 2 A_0
\end{pmatrix} \psi \Big \|_{\mathcal{H}(\underline{A})}=  2 \left\|  A_0 \psi_2 \right\|  \leq 2 \,\tilde{\delta}^{1/2}
\Big \|    \begin{pmatrix}
0  & i \\ -i B(\underline A)^2 & 0
\end{pmatrix} \psi  \Big  \|_{\mathcal{H}(\underline{A})} + 
2 \, \left( C_{\tilde{\delta}}\right)^{1/2}\| \psi \|_{\mathcal{H}(\underline{A})},  
\end{align} 
thus
 $ 
  \begin{pmatrix}
0 & 0 \\ 0 & 2 A_0
\end{pmatrix}  
$
 is small in the sense of Kato with respect to
$   \begin{pmatrix}
0  & i \\ -i B(\underline A)^2 & 0
\end{pmatrix} 
$, with relative bound smaller than $1$ (here we use that $\delta < \frac{1}{5}$).  The fact that $H(\underline A)$ is self-adjoint follows from this last assertion and the Kato-Rellich theorem (Theorem X.12 in \cite{rs2}).

\bull

\subsection{The Wave and Scattering Operators}\label{wave-scattering}

\subsubsection{Wave Operators} \label{exwave}
In this paragraph we prove that the wave operators exist. Lemma \ref{Lexsh} proves the existence of the wave operators in the case that $A \in \mathcal{A}^{({\bf reg})}_\Phi(B) $ (the regular case). The general case is proved in Section \ref{existence-wave}. In Section \ref{existence-scattering} we study the scattering operator. Related results are presented in \cite{w3}.   
 We identify $\kappa$ (see \eqref{2.4}) with the multiplication operator by the matrix
\beq \label{mkappa}
\begin{pmatrix}
\kappa & 0 \\ 0 & \kappa \end{pmatrix} \equiv \kappa.
\ene
\begin{lemma} \label{waveo}
For every $A\in \mathcal{A}_{\Phi}(B)$,  the limits \eqref{2.19} exist, if and only if, the limits 
$  \hbox{\rm s-}\lim_{t \rightarrow \pm \infty} e^{it
H(\underline A)}\, \kappa \, e^{-it H_0} $ exist. In either case 
\beq 
  W_{\pm}(\underline A) = \hbox{\rm s-}\lim_{t \rightarrow \pm \infty} e^{it
H(\underline A)}\, \kappa \, e^{-it H_0}.
\label{waveos}
\ene
\end{lemma}
\noindent \emph{Proof:} It is enough to prove that for every $\phi  \in \mathcal{S}(\mathbb{R}^3) \oplus \mathcal{S}(\mathbb{R}^3)$
\beq \label{waveo1}
\lim_{t \to \pm \infty}\Big \| \Big [ \begin{pmatrix}
 B(\underline A)^{-1} \chi_\Lambda B_0 & 0 \\ 0 & \chi_\Lambda 
\end{pmatrix} - \kappa \Big ] e^{- it H_0} \phi \Big \|_{\mathcal H(\underline A)} = 0.   
\ene 
As $B_0$ has only absolutely continuous spectrum, $e^{i t B_0}$ converges weakly to zero when 
$t$ tends to $\pm \infty$ (using the Riemann-Lebesgue Lemma). The Rellich-Kondrachov theorem
implies that
\beq  \label{rk}
\lim_{t \rightarrow \pm \infty}\left\|  f(x)\, e^{it B_0}\, \phi \right\|=0, \forall \phi \in L^2\left(\mathbb R^3\right), \,    \mathrm{if}\,
\lim_{r \rightarrow \infty }|f(r \hat{x})|=0,\, \mathrm{uniformly \, on} \, \hat{x} \in \ese^2. 
\ene
 From this and \eqref{diagu} we deduce that \eqref{waveo1} 
 is fulfilled whenever
\beq \label{waveo2}
\lim_{t \to \pm \infty} \|  (\kappa B_0  - B(\underline A)\kappa) e^{- i t B_0} \psi \| =
\lim_{t \to \pm \infty} \| e^{it B(\underline A)} (\kappa B_0  - B(\underline A)\kappa) e^{- i t B_0}  \psi \| = 0, 
\ene
for every $\psi \in \mathcal{S}(\mathbb{R}^3)$.
We complete our proof showing the validity of \eqref{waveo2}.\\

It follows from the proof of Lemma  5.3 in \cite{bw} that the limits 
\beq \label{waveo3}
W_{\pm}(B(\underline A)^2; B_0^2 ) : = s-\lim_{t \to \pm \infty} e^{it B(\underline A)^2} \kappa 
e^{-i B_0^2} 
\ene
exist (they are actually the wave operators for the Schr\"odinger equation). The invariance principle (Theorem XI.23 in \cite{rs3}) implies that the limits 
\beq \label{waveo4}
W_{\pm}(B(\underline A); B_0 ) : = s-\lim_{t \to \pm \infty} e^{it B(\underline A)} \kappa 
e^{-i B_0}  
\ene
exist and are equal to \eqref{waveo3}. Actually  \cite{rs3} does not consider obstacles, however, the proof also applies in this case. It follows that 
\beq \label{waveo5}
\lim_{t \to \pm \infty}  e^{it B(\underline A)} \kappa B_0  e^{- i t B_0}  \psi 
= W_{\pm}(B(\underline A); B_0 ) B_0 \psi.
\ene
Now we analyze the term $  e^{it B(\underline A)}  B(\underline A)\kappa  e^{- i t B_0}  \psi  $. We prove that 
\beq \label{waveo6}
\lim_{t \to \pm \infty} \|  B(\underline A)  e^{it B(\underline A)} \kappa  e^{- i t B_0}  \psi \|
= \| W_{\pm}(B(\underline A); B_0 ) B_0 \psi \|.
\ene
To obtain \eqref{waveo6} we compute [see \eqref{2.15} and \eqref{domq}]
\begin{align} \label{waveo7}
 \Big | \|  B(A)   \kappa  e^{- i t B_0}  \psi \|^2 - \|     \kappa  e^{- i t B_0} B_0 \psi \|^2 \Big |
= &  
 \Big | \Big \langle (\mo - A)   \kappa  e^{- i t B_0}  \psi, \, 
(\mo - A)   \kappa  e^{- i t B_0}  \psi \Big \rangle 
\\  & \notag  + 
m^2 \Big \langle   \kappa  e^{- i t B_0}  \psi, \, 
   \kappa  e^{- i t B_0}  \psi \Big \rangle
   - \Big \langle A_0^2   \kappa  e^{- i t B_0}  \psi, \, 
   \kappa  e^{- i t B_0}  \psi \Big \rangle  -  \|     \kappa  e^{- i t B_0} B_0 \psi \|^2 
   \Big |
  \\  & \notag  \longrightarrow 0,  \hspace{2cm} \text{as} \: \: t \to \pm \infty,
\end{align}
 where we use  \eqref{rk}. Notice that we can substitute the last term  $  \|     \kappa  e^{- i t B_0} B_0 \psi \|^2   $ by $   \|     e^{- i t B_0} B_0 \psi \|^2 = 
  \| \mo      e^{- i t B_0} \psi \|^2 +  m^2 \|      e^{- i t B_0} \psi \|^2  $, using   \eqref{rk}, and we can handle similarly the other terms containing $\kappa$. \\
With the help of \eqref{waveo5} and \eqref{waveo6} we obtain
\begin{align} \label{waveo8}
\lim_{t \to \pm \infty} 
  \| e^{it B(\underline A)} (\kappa B_0 & - B(\underline A)\kappa ) e^{- i t B_0}  \psi \|^2 \\ \notag  = & \lim_{t \to \pm \infty} 
\Big [ 2 \| W_{\pm}(B(\underline A); B_0 ) B_0 \psi\|^2
   - \Big \langle e^{it B(\underline A)} B(\underline A)\kappa e^{- i t B_0}  \psi,   
W_{\pm}(B(\underline A); B_0 ) B_0 \psi 
 \Big  \rangle 
\\ & \notag \hspace{4.5cm}  - 
 \Big \langle   
W_{\pm}(B(\underline A); B_0 ) B_0 \psi,  e^{it B(\underline A)} B(\underline A)\kappa e^{- i t B_0}  \psi 
 \Big  \rangle \Big ]  = 0, 
\end{align}
where we used that
\begin{align}
\lim_{t \to \pm \infty} \Big \langle e^{it B(\underline A)} B(\underline A)\kappa e^{- i t B_0}  \psi,   
W_{\pm}(B(\underline A); B_0 ) B_0 \psi 
 \Big  \rangle  = & \lim_{t \to \pm \infty} \Big \langle e^{it B(\underline A)} \kappa e^{- i t B_0}  \psi,   
W_{\pm}(B(\underline A); B_0 ) B_0^2 \psi 
 \Big  \rangle \\ \notag = &  \| W_{\pm}(B(\underline A); B_0 ) B_0 \psi\|^2.
\end{align}
\bull

\begin{lemma}[Existence: The Regular Case] \label{Lexsh}
For every $A \in \mathcal{A}^{({\rm reg})}_\Phi(B)$ 
 the limits \eqref{waveos} exist and are isometric. 
\end{lemma}
\noindent \emph{Proof:} The result follows from standard techniques using Cooks's argument, \eqref{diagu} and Lemma \ref{pento}: For $\phi = (\phi_1, \phi_2) \in \mathcal{S}(\mathbb{R}^3)^2$, with $\widehat \phi_1, \widehat \phi_2$ satisfying the properties of $f$ in Lemma \ref{pento} (note that these functions are dense in $L^2\left ( \mathbb R^3 \right)^2$) , we write 
\beq \label{este1}
e^{i t H(\underline A)} \kappa e^{-it H_0} \phi = 
\kappa \phi + \int_{0}^s e^{i t H(\underline A)} \big [H(\underline A)\kappa - \kappa H_0 \big]  e^{-is H_0} \phi .
\ene
Set $ h(s) = \Big \|   \big [H(\underline A)\kappa - \kappa H_0 \big]  e^{-is H_0} \phi  \Big \|_{\mathcal{H}(\underline A)} $. We bound $h$ by an integrable function using \eqref{diagu} and Corollary  \ref{cocal}. 
This allows us to take the limit $t \to \pm \infty$ in \eqref{este1}. The isometric property 
follows from \eqref{J}, \eqref{2.19} and \eqref{rk}.

\subsubsection{Existence of the Wave Operators (the General Case)}\label{existence-wave}
In this section we prove existence of wave operators \eqref{waveos} for every magnetic potential $A \in \mathcal{A}_\Phi(B)$ using Lemma 
\ref{Lexsh} and a change of gauge argument. We provide additionally a change of gauge formula.

\begin{theorem}[Existence of Wave Operators and Change of Gauge Formula] \label{exchan}
For every $ \underline A = (A_0, A)$  with  $ A \in A_{\Phi}(B)$  the limits \eqref{waveos} exist and are isometric. For every 
$ \underline A^{(i)} = (A_0, A^{(i)})$, $i \in \{1, 2 \}$, with 
$ A^{(1)},  A^{(2)} \in  \mathcal{A}_{\phi}(B)$
\beq \label{changeW}
W_{\pm}(\underline A^{(2)}) = e^{ i \lambda(x) }W_{\pm}(\underline A^{(1)}) e^{- i \lambda_\infty },
\ene
 where $A^{(2)} - A^{(1)} = \nabla \lambda $ [see \eqref{2.11}-\eqref{lambdainfty}]. 

\end{theorem}
\noindent \emph{Proof:}
We suppose that $A^{(1)} \in \mathcal{A}^{({\bf reg})}(B)$. Lemma \ref{Lexsh} assures the existence of the wave operators
$  W_{\pm}(\underline A^{(1)}) $. By proving the change of gauge formula we prove the existence of $  W_{\pm}(\underline A^{(2)}) $. The same proof applies for general $A^{(1)}$ and $A^{(2)}$. We prove the assertion for $W_{+}$. The proof for $ W_{-}$ is analogous. A simple computation gives 
\beq \label{PRW1} 
H(\underline A^{(2)} ) =  e^{i \lambda (x)} H(\underline A^{(1)} )e^{-i \lambda (x)}, 
\ene 
which implies that
\beq \label{PRW2}
W_+(\underline A^{(2)}) = e^{i \lambda (x)} s- \lim_{t \to \infty} e^{it H(\underline A^{(1)})} \kappa e^{-it H_0} 
e^{it H_0} \kappa' e^{- i \lambda(x)} e^{-it H_0},   
\ene
whenever the limit exists. Here $\kappa' $ satisfies the properties of $\kappa$ (see \eqref{2.4} and the text above)  and \eqref{mkappa}. Additionally
\beq \label{PRW3}
\kappa' \kappa = \kappa.
\ene
Then we only need to prove that
\beq \label{PRW4}
 s-\lim_{t \to \infty} 
e^{it H_0} \kappa' e^{- i \lambda(x)} e^{-it H_0} =  e^{-i\lambda_\infty}. 
\ene
Using \eqref{fw}-\eqref{diagu} we obtain that  
\beq \label{PRW5}
e^{it H_0} \kappa' e^{- i \lambda(x)} e^{-it H_0} = U^{-1 }  \begin{pmatrix}
e^{it B_0} & 0 \\ 0 & e^{-it B_0}
\end{pmatrix} Q  \begin{pmatrix}
 B_0 \kappa' e^{- i \lambda(x)}B_0^{-1} & 0 \\ 0 &  \kappa' e^{- i \lambda(x)}
\end{pmatrix}     Q^{-1 }  \begin{pmatrix}
e^{-it B_0} & 0 \\ 0 & e^{it B_0}
\end{pmatrix} U .
\ene
By \eqref{2.12}, \eqref{rk} 
\beq \label{PRW7}
s-\lim_{t\to \infty}   \big [ e^{- i \lambda_\infty } -   e^{- i \lambda(x)} \big ]  e^{\pm i t B_0}  = 0. 
\ene
Then by  \eqref{PRW5},   \eqref{PRW7},  
\beq \label{PRW6}
s-\lim_{ t \to \infty} \Big[
e^{it H_0} \kappa' e^{- i \lambda(x)} e^{-it H_0} - U^{-1 }  \begin{pmatrix}
e^{it B_0} & 0 \\ 0 & e^{-it B_0}
\end{pmatrix}  \begin{pmatrix}
  e^{- i \lambda_\infty} & 0 \\ 0 &  e^{- i \lambda_\infty}
\end{pmatrix}      \begin{pmatrix}
e^{-it B_0} & 0 \\ 0 & e^{it B_0}
\end{pmatrix} U \Big] = 0.
\ene
We recall that $\lambda_\infty $ is a constant function (see the text below Remark \ref{Rreg}) to prove \eqref{PRW4}. To deduce \eqref{PRW6} we use that 
\begin{eqnarray*} \label{PRW7prima}
\| B_0  \big [ e^{- i \lambda_\infty } -  \kappa'\, e^{- i \lambda(x)} \big ]  B_0^{-1}
e^{\pm i t B_0} \|^2 =  \| \mo  \big [ e^{- i \lambda_\infty } -  \kappa'\, e^{- i \lambda(x)} \big ]  B_0^{-1}
e^{\pm i t B_0} \|^2  + \\ m^2 \|   \big [ e^{- i \lambda_\infty } - \kappa'\,  e^{- i \lambda(x)} \big ]  B_0^{-1}
e^{\pm i t B_0} \|^2
\rightarrow 0,\, \mathrm{as}\, t \rightarrow \pm \infty,
\end{eqnarray*}
and that the terms containing derivatives of $ \kappa  e^{- i \lambda(x)} $ vanish as 
$t$ tends to infinity, using  \eqref{rk}.     

 \subsubsection{Scattering Operator} \label{existence-scattering}

The following theorem proves that the scattering operator is invariant under change of gauge; it is a direct consequence of Theorem \ref{exchan}.  
\begin{theorem}\label{changescat}
For every 
$ \underline A^{(i)} = (A_0, A^{(i)}), i =1, 2$, with 
$ A^{(1)},  A^{(2)} \in  \mathcal{A}_{\phi}(B)$,
\beq \label{change}
S(\underline A^{(2)}) =  S(\underline A^{(1)}). 
\ene
\end{theorem}

\begin{remark}{\rm As in \cite{bw} we can  consider change of gauge formulae when $ A^{(1)},  A^{(2)}$ only have the same fluxes modulo $2 \pi$. We do not dwell on these issues here.
}
\end{remark}

\subsection{High Momenta Limit of the Scattering Operator}\label{main}
In this section we prove our main results: We give a high-momenta expression for the scattering operator, with error bounds. This formula is the content of Theorem \ref{TPG}, whose proof in the main purpose of Section \ref{high-long}. To prove our theorem we first prove the result in the special case that $A$ is regular ($A \in \mathcal{A}^{{(\rm reg)}}_\Phi(B)$). This is the main result of Section \ref{high-short}. The general result follows from a change of gauge formula, which is accomplished  in Section \ref{high-long}. Our formula is used in Sections
\ref{Sinv} and \ref{San} to reconstruct important information from the potentials and the magnetic field.

\subsubsection{High Momenta  Limit for the Scattering Operator: The Regular Case}\label{high-short}
This section is the most laborious part of our paper. Here we estimate the high-momenta limit of the scattering operator in the case that  $A$ is regular ($A \in \mathcal{A}^{{(\rm reg)}}_\Phi(B)$). This is actually a relevant result, since the general case follows from it by a simple gauge-argument. Our main result in this section is Theorem \ref{TP}. The whole section is devoted to prove preliminary lemmata and theorems that we use  in the proof of Theorem \ref{TP}. To analyze the scattering operator, the wave operators are fundamental. As we can see from \eqref{waveos}, the wave operators represent asymptotic limits when the time goes to plus or minus infinity. To do our proof we analyze $e^{it H(\underline A)}\kappa e^{-itH_0}$ for finite time $t$, in the high momenta regime, and then we bound the error $ W_{\pm }(\underline A) - e^{it H(\underline A)}\kappa e^{-itH_0} $. The high-momenta analysis of $e^{it H(\underline A)}\kappa e^{-itH_0}$ is the most complicated part, it is done in Theorem \ref{T:e0}, which is based in Lemmata \ref{mero} and \ref{meroli}. Lemmata \ref{meropp} and \ref{nosta} deal with the error $ W_{\pm }(\underline A) - e^{it H(\underline A)}\kappa e^{-itH_0} $. These lemmata together with Theorem \ref{T:e0} and Lemma \ref{L:i1} (which is a technical result concerning the norms we are using) are the ingredients we need to prove our main result (Theorem \ref{TP}).

%
%

\begin{definition} {\rm
Let $ \underline A = (A_0, A)$  with  $ A \in A_{\Phi}(B)$.  
For every $v > 0$ and every $\nu \in \mathbb{S}^2$
we define (see \eqref{2.4} and the text above it)
\begin{align}\label{a1}
Z_{\underline A}(t,  v\nu , \kappa )= e^{i t \beta( v + \nu\cdot \mo )}
  \begin{pmatrix}  0 & 0 \\ i  2 \kappa A \cdot \nu  &  2\kappa A_{0} \end{pmatrix}
e^{-i t \beta( v + \nu \cdot \mo )} 
\end{align} }
\end{definition}

\begin{lemma} \label{mero}
Let $\nu \in  \mathbb{S}^2 $ and $l \in \mathbb{N}$,  $l \geq \zeta/2$.
Suppose that $ \phi \in  {\bf H}_{\langle x \rangle^{4 l}}^\alpha(\mathbb{R}^3)^2$, $\alpha \geq 2 $, is supported in $\Lambda_{\kappa, \nu} $ (see \eqref{Lambdanu}).
Let $ \underline A = (A_0, A)$,  with  $ A \in A_{\Phi}^{({\rm reg})}(B)$.
The for every $v \geq v_0$ (see Lemma \ref{pento}).

\begin{align}\label{e21}
 \int_{0}^{ t}e^{is H(\underline A)} i (H(\underline A) \kappa - \kappa H_{0} ) e^{-is H_{0}} F_{W}^{-1} e^{ix \cdot v \nu}
 \phi \, ds
= & 
 \int_{0}^{ t}e^{is H(\underline A)} \kappa \,e^{-is H_{0}} F_{W}^{-1} e^{ix \cdot v \nu}i Z_{\underline A}(t,  v\nu , \kappa ) 
 \phi \, ds \notag \\
& + O\Big(  \int_0^{ |t|} ds 
  \Big [  \|    \phi   \|_{{\bf H}_{\langle x \rangle^{4 l}}^\alpha(\mathbb{R}^3)^2}\Big(  \frac{1}{v ( 1 + |s|)^{2l - 1}}  + 
\sum_{j = 1}^{2l}\frac{1}{(1 + |s|)^{2l - j} v^{\min(\alpha, j)}} \Big) \notag \\  & \hspace{3cm}  + \| \phi \|_{{\bf H}_{\langle x \rangle^{4 l}}^2(\mathbb{R}^3)^2} \frac{1}{v ( 1 + |s|)^{\zeta - 1}}  \Big ]  + \frac{|t|}{v^\alpha} \| \phi \|_{{\bf H}^{\alpha}(\mathbb{R}^3)^2}   \Big), 
\end{align}
for every $t\in \mathbb{R} $.

\end{lemma}
\noindent \emph{Proof:}
First we compute
\begin{align} \label{e21e}
 \int_{0}^{t} 
e^{is H(\underline A)} i ( H(\underline A)\kappa - & \kappa H_{0} )  e^{-isH_{0} } F_{W}^{-1} e^{i x\cdot v \nu} 
\phi \\ = &   \notag 
 \int_{0}^{t} e^{is H(\underline A)}  e^{i x \cdot v \nu } e^{-i x \cdot v \nu } 
 i( H(\underline A)\kappa - 
\kappa H_{0} )
 F_{W}^{-1} e^{i x \cdot v \nu } e^{-i x \cdot v \nu}
 F_{W} e^{-isH_{0} } F_{W}^{-1} e^{i x\cdot v \nu} \phi. \notag  
\end{align}  
It follows from \eqref{fw1} that 
\begin{align}\label{a6}
F_{W} e^{ \pm i t H_{0}} F_{W}^{-1} = e^{\pm i t \beta (\mo^{2}+ m ^{2})^{1/2}}.  
\end{align}
Notice that
\begin{align}\label{i6}
& (H(A) \kappa - \kappa H_{0})F_{W}^{-1} = \begin{pmatrix}  0 & 0 \\ \vartheta \frac{1}{(\mo^{2}+ m^{2})^{1/2}} & 2\kappa A_{0} \end{pmatrix},  
\end{align}  
where
\begin{align}\label{vartheta} 
\vartheta = -i((\mo^{2}\kappa)   + 2 ((\mo \kappa) \cdot \mo) - 
(\mo\cdot A) \kappa- 2 A \cdot (\mo \kappa) - 2 \kappa A \cdot \mo + \kappa A^{2} - \kappa A_{0}^{2} ). 
\end{align}
From (\ref{usefullEq}) , (\ref{i6}) and the fact that $ (H(\underline A) \kappa - \kappa H_{0})F_{W}^{-1} $ 
have only zeros in the first row follows
\begin{align}\label{a8}
\| e^{-i x \cdot v \nu} (H(\underline A) \kappa - \kappa H_{0})F_{W}^{-1} e^{ i x \cdot v \nu } 
\|_{{\cal L}(L^{2}(\mathbb{R}^{n})\oplus L^{2}(\mathbb{R}^{n}), {\cal H}(\underline A))} \leqq C,
\end{align}
where $ C $ is a constant independent of  $ v \nu $. \\
We have that
\begin{align}\label{a9}
& e ^{i x \cdot v \nu } e^{-i x \cdot v \nu} (H(\underline A) \kappa - \kappa H_{0})F_{W}^{-1} e^{ i x \cdot v \nu } = 
 \begin{pmatrix}  0  & 0 \\ 0 & e^{i x \cdot v \nu}  \end{pmatrix} e^{-i x \cdot v \nu} (H(\underline A) \kappa - \kappa H_{0})F_{W}^{-1} 
e^{ i x \cdot v \nu }, 
\end{align}
and clearly 
\beq \label{cle}
 \Big \|  \begin{pmatrix}  0  & 0 \\ 0 & e^{i x \cdot v \nu }  \end{pmatrix}  
\Big \|_{{\cal L}({\cal H} (\underline A), {\cal H}(\underline A))} = 1.
\ene 
From Lemmata \ref{pen1}, \ref{estep} and from Eqs.  \eqref{2.4}, \eqref{reg}, \eqref{e21e}-\eqref{cle}, it follows that (see also \eqref{cansa3} and see \eqref{B0}, \eqref{fw1} and  \eqref{usefullEq} )
\begin{align}\label{a10} 
 \int_{0}^{t} 
e^{is H(\underline A)} i ( H(\underline A)\kappa - & \kappa H_{0} )  e^{-isH_{0} } F_{W}^{-1} e^{i x\cdot v \nu} 
\phi  \\  \notag 
 = & \,
  \int_{0}^{t} e^{is H(\underline A)}  e^{i x \cdot v \nu } i 
\begin{pmatrix} 0 & 0 \\ -i ( 2 (\mo \kappa) \cdot \nu - 2 \kappa A \cdot \nu ) & 2 \kappa A_{0}  \end{pmatrix}
e^{-i s\beta ( v + \nu \cdot \mo )}  \phi \\
& + O\Big(  \int_0^{ |t|} ds 
  \Big [  \|    \phi   \|_{{\bf H}_{\langle x \rangle^{4 l}}^\alpha(\mathbb{R}^3)^2}\Big(  \frac{1}{v ( 1 + |s|)^{2l -1}}  + 
\sum_{j = 1}^{2l}\frac{1}{(1 + |s|)^{2l - j} v^{\min(\alpha, j)}} \Big) \notag \\  & \hspace{6cm}  + \| \phi \|_{{\bf H}_{\langle x \rangle^{4 l}}^2(\mathbb{R}^3)^2} \frac{1}{v ( 1 + |s|)^{\zeta - 1}}  \Big ]  + \frac{|t|}{v^\alpha} \| \phi \|_{{\bf H}^{\alpha}(\mathbb{R}^3)^2}   \Big). \notag 
\end{align} 
Since $ \phi $ is supported in $\Lambda_{\kappa, \nu} $ (see \eqref{Lambdanu}), and using that $e ^{i s \nu \cdot \mo}$ is a translation operator in $L^2(\mathbb{R}^2)$, we prove that
\begin{equation} \label{a11}
\begin{pmatrix} 0 & 0 \\ -i ( 2 (\mo \kappa) \cdot \nu - 2 \kappa A \cdot \nu ) & 2 \kappa A_{0}  \end{pmatrix}
e^{-i s\beta ( v + \nu \cdot \mo )}  \phi = 
\kappa \begin{pmatrix} 0 & 0 \\ 2 i \kappa A \cdot \nu  & 2 \kappa A_{0}  \end{pmatrix}
e^{-i s\beta ( v + \nu \cdot \mo )}  \phi.
\end{equation}
Now we consider that, see \eqref{B0}, \eqref{fw1} 
\begin{align} \label{noeq}
 e^{is H(\underline A)}  e^{i x \cdot v \nu } \kappa i 
\begin{pmatrix} 0 & 0 \\ i  2 \kappa A \cdot \nu  & 2 \kappa A_{0}  \end{pmatrix}
& = e^{is H(\underline A)}  \kappa  e^{-i sH_0} F_W^{-1}   e^{i x \cdot v \nu } 
  e^{- i x \cdot v \nu }  F_W e^{-i sH_0}  F_W^{-1} e^{i x \cdot v \nu }   i 
\begin{pmatrix} 0 & 0 \\ i  2 \kappa A \cdot \nu  & 2 \kappa A_{0}  \end{pmatrix}
\\ \notag  & = e^{is H(\underline A)}  \kappa  e^{-i sH_0} F_W^{-1}   e^{i x \cdot v \nu } 
  \big [ e^{- i x \cdot v \nu }  e^{-i s B_0 \beta }   e^{i x \cdot v \nu } \big ]  i 
\begin{pmatrix} 0 & 0 \\ i  2 \kappa A \cdot \nu  & 2 \kappa A_{0}  \end{pmatrix},
\end{align}
where we use that $ F_W^{-1} \begin{pmatrix}
0 & 0 \\ X & Y
\end{pmatrix} = \begin{pmatrix}
0 & 0 \\ X & Y
\end{pmatrix} $, for every matrix of the form $ \begin{pmatrix}
0 & 0 \\ X & Y
\end{pmatrix} $. We additionally use that $F_W^{-1} : L^2(\mathbb{R}^3)^2 \to \mathcal{H}_0 $ is unitary and that $\kappa : \mathcal{H}_0 \to
\mathcal{H}(\underline{A})$ is bounded (actually $ \kappa  $ is bounded from ${\bf H}^1(\mathbb{R}^3) $ with values in $ {\bf H}^1(\Lambda) $).      
Eqs. \eqref{2.4}, \eqref{reg}, \eqref{a10}-\eqref{noeq}, \eqref{a5}, \eqref{usefullEq} and Lemma \ref{pent1}  imply, see also \eqref{cansa3}, \eqref{e21}. 

\begin{lemma} \label{meropp}
Let $\nu \in  \mathbb{S}^2 $ and $l \in \mathbb{N}$,  $l \geq \zeta/2$.
Suppose that $ \phi \in  {\bf H}_{\langle x \rangle^{4 l}}^0(\mathbb{R}^3)^2 \cap 
  {\bf H}^\alpha(\mathbb{R}^3)^2 $, $\alpha > 0 $, is supported in $\Lambda_{\kappa, \nu} $. Let $ \underline A = (A_0, A)$,  with  $ A \in A_{\Phi}^{({\rm reg})}(B)$.
Then, for every $v \geq v_0$ (see Lemma \ref{pento}),
\begin{align} \label{pg1pp}
\Big \|\Big [ W_{\pm}(\underline A)   -  e^{it H(\underline A)} \kappa e^{-itH_{0} } \Big ] F_{W}^{-1} e^{i x\cdot v \nu}\phi  \Big  \|_{\mathcal{H}(\underline A)}
\leq  C  \|    \phi   \|_{{\bf H}_{\langle x \rangle^{4 l}}^0(\mathbb{R}^3)^2}
\Big ( \frac{1}{1 + |t|} \Big )^{\zeta -1} 
+ C \|    \phi   \|_{{\bf H}^\alpha(\mathbb{R}^3)^2}\frac{1}{v^\alpha},
\end{align}
for every $t\in \mathbb{R} $, such that $\pm t \geq 0$. 
\end{lemma}
\noindent \emph{Proof:} We take, without loss of generality, the plus sign in $W_{\pm}(\underline A)$. 
Set $\tau \in C_0^\infty(\mathbb{R}^3; [0, 1])$ be such that 
$\tau(x) = 1$ for $|x| \leq \frac{1}{2} $ and it vanishes for $ |x| \geq 1 $. 
By Duhamel's formula
\begin{align} \label{e21epp}
\Big [ W_{+}(\underline A)   -  e^{it H(\underline A)} \kappa e^{-itH_{0} } \Big ] F_{W}^{-1} e^{i x\cdot v \nu}  \tau(16 \mo/v) \phi  = 
&  \int_{t}^{\infty} 
e^{is H(\underline A)} i ( H(\underline A)\kappa - \kappa H_{0} )  e^{-isH_{0} } F_{W}^{-1} e^{i x\cdot v \nu} \tau(16 \mo/v)
\phi \\ = &   \notag   
 \int_{t}^{\infty} e^{is H(\underline A)}  e^{i x \cdot v \nu } e^{-i x \cdot v \nu } 
 i( H(\underline A)\kappa - 
\kappa H_{0} ) \\ & \notag \hspace{2.8cm} 
 F_{W}^{-1} e^{i x \cdot v \nu } e^{-i x \cdot v \nu}
 F_{W} e^{-isH_{0} } F_{W}^{-1} e^{i x\cdot v \nu}  \tau(16 \mo/v) \phi . \notag  
\end{align}  
Using \eqref{a6}-\eqref{cle}, it follows from Lemma \ref{pen1o} (see also
 \eqref{usefullEq}, \eqref{2.4} and \eqref{reg}) that for every $ l \in \mathbb{N}$ there is a constant $C_{l}$ such that
\begin{align} \label{he1ol}
\Big \| e^{-i x \cdot v \nu } 
 i( H(\underline A)\kappa - 
\kappa H_{0} ) 
 F_{W}^{-1} e^{i x \cdot v \nu } e^{-i x \cdot v \nu}
 F_{W} e^{-isH_{0} }  &  F_{W}^{-1} e^{i x\cdot v \nu}  \tau(16 \mo/v) \phi \Big \|_{\mathcal{H}(\underline A)}  \\ \notag  &  \leq 
C_{l}  \Big [  \|    \phi   \|_{{\bf H}_{\langle x \rangle^{4 l}}^0(\mathbb{R}^3)}\Big(  \frac{1}{( 1 + |t|)^{2l}}  \Big) 
 + \| \phi \|_{{\bf H}^0(\mathbb{R}^3)} \frac{1}{ ( 1 + |t|)^{\zeta }}  \Big ], 
\end{align}
for every $v > v_0$ (see Lemma \ref{pento}), $\nu \in \mathbb{S}^2$, $t \geq 0$, and every $\phi \in {\bf H}_{\langle x \rangle^{4 l}}^0(\mathbb{R}^3) $. Eq. \eqref{he1ol} together with \eqref{cansa3}, \eqref{e21epp} and \eqref{cle} imply the desired result.

\bull

\begin{lemma} \label{meroli}
Take $\nu \in  \mathbb{S}^2 $ and 
suppose that $ \phi \in  {\bf H}_{\langle x \rangle^{2}}^0(\mathbb{R}^3)^2$. Let $ \underline A = (A_0, A)$,  with  $ A \in A_{\Phi}^{({\rm reg})}(B)$.
Then, for every $v \geq v_0$ (see Lemma \ref{pento}) and every $t \in  \mathbb{R} $ 
\begin{align}\label{e21mta} 
 \int_{0}^{ t}e^{is H(\underline A)} \kappa e^{-is H_{0}} F_{W}^{-1} e^{ix \cdot v \nu}i Z_{\underline A}(t,  v\nu , \kappa ) \,
\, \phi\, ds 
=  
\int_{0}^{ t}e^{is H(\underline A)} \kappa e^{-is H_{0}} F_{W}^{-1} e^{ix \cdot v \nu}i W_{\underline A}(t,  \nu , \kappa )  \, \phi \, ds + O\Big ( \frac{1}{v}\Big )\| \phi 
\|_{ {\bf H}_{\langle x \rangle^{4}}^0(\mathbb{R}^3)^2},  
\end{align}
recall Definition \ref{WUA}. 
\end{lemma}
\noindent \emph{Proof:} We define 
\begin{align}\label{e30}
Z'=  \begin{pmatrix}  - \kappa A_{0} & i \kappa A \cdot \nu  \\ i \kappa A \cdot \nu  & \kappa A_{0} \end{pmatrix},  & &
W' =  \begin{pmatrix}   \kappa A_{0} & -i \kappa A \cdot \nu  \\  i \kappa A \cdot \nu  & 
\kappa A_{0} \end{pmatrix} 
\end{align}
as  $ Z' $ anti-commutes with $ \beta $ and $ W' $ commutes with $ \beta $, 
\begin{align}\label{e31}
Z_{\underline A}(s, v \nu , \kappa ) \phi = 
(W_{\underline A}(s, \nu, \kappa) + e^{2i s \beta v} e^{i s \beta \nu \cdot \mo} Z' e^{-i s\beta \nu \cdot \mo } )\phi . 
\end{align} 
We plug \eqref{e31} in the left-hand side of \eqref{e21mta} and estimate the term corresponding to the second term in \eqref{e31} using an integration by parts: 
\begin{align}\label{e32}
 \int_{0}^{ t}e^{is H(\underline A)} \kappa  e^{-is H_{0}} F_{W}^{-1} & 
e^{ix \cdot v \nu } \ i \ e^{2i s \beta v} e^{i s\beta \nu \cdot \mo} Z' e^{-i s\beta \nu \cdot \mo } 
 \phi   \\
= &   e^{is H(\underline A)} \kappa e^{-is H_{0}} F_{W}^{-1} 
e^{ix \cdot v\nu}i(2i \beta v)^{-1}  e^{2i s \beta v} e^{i s\beta \nu \cdot \mo } Z' e^{-i s\beta \nu \cdot \mo } 
 \phi \bigg|_{0}^{ t} \notag   \\
& - \int_{0}^{ t}( \frac{\partial}{\partial s}e^{is H(\underline A)} \kappa e^{-is H_{0}}) F_{W}^{-1} 
e^{ix \cdot v \nu }i (2i \beta v)^{-1} e^{2i s \beta v} e^{i s\beta \nu \cdot \mo} Z' e^{-i s\beta \nu \cdot \mo } 
 \phi \notag  \\
& - \int_{0}^{ t}e^{is H(\underline A)} \kappa e^{-is H_{0}} F_{W}^{-1} 
e^{ix \cdot v \nu }(2i \beta v)^{-1} i e^{2i s \beta v} ( \frac{\partial}{\partial s} e^{i s\beta \nu \cdot \mo} Z' e^{-i s\beta \nu \cdot \mo } ) 
 \phi  \notag \\ 
= & \: O\Big (\frac{1}{v} \Big ) \| \phi 
\|_{ {\bf H}_{\langle x \rangle^{4}}^0(\mathbb{R}^3)^2} \notag.
\end{align} 
To justify the last line in Eq. (\ref{e32}), we notice that following the procedure 
of the proof of Corollary \ref{cocal}, using Remark \ref{rpento}, we prove that
there is an integrable function $ h(s) $ such that 
$ \|  Z' e^{-i s\beta \nu \cdot \mo }  $  $
 \phi  \|_{(L^{2}(\mathbb{R}^{n}))^{2}} $  $+ \sum_{i = 1}^{3} $  $ \| 
 (\frac{\partial }{\partial x_{i}}  Z') $  $ e^{-i s\beta \nu \cdot \mo }  $  $ 
 \phi  \|_{(L^{2}(\mathbb{R}^{n}))^{2}} $ $  \leqq  h(s) \| \phi 
\|_{ {\bf H}_{\langle x \rangle^{4}}^0(\mathbb{R}^3)^2}   $.
We additionally notice that $ ( \frac{\partial}{\partial s}  e^{is H(\underline A)} $ $ \kappa e^{-is H_{0}}) $ $ F_{W}^{-1} $ $ 
e^{ix \cdot v\nu } $ $ : L^{2}(\mathbb{R}^{3})^{2} \to {\cal H}(\underline  A) $ is bounded, independently of
$ s  $ and $ v \nu $. Eqs. \eqref{e30}-\eqref{e32} imply \eqref{e21mta}.  
\bull

\begin{theorem}\label{T:e0}
Set $\nu \in  \mathbb{S}^2 $ and $l \in \mathbb{N}$,  $l \geq \zeta/2$.
Suppose that $ \phi \in  {\bf H}_{\langle x \rangle^{4 l}}^2(\mathbb{R}^3)^2$
 is supported in $\Lambda_{\kappa, \nu} $. Let $ \underline A = (A_0, A)$,  with  $ A \in A_{\Phi}^{({\rm reg})}(B)$.
Then, for every $v \geq v_0$ (see Lemma \ref{pento}) and every $t \in  \mathbb{R} $ 
\begin{align}\label{e33}
 e^{i t H(\underline A)} \kappa e^{-i t H_{0}} F_{W}^{-1}e^{i x \cdot v \nu}  \phi 
= & \kappa F_{W}^{-1} e^{ix \cdot v \nu} e^{i \int_{0}^{t} W_{\underline A}(r, \nu, \kappa) dr } \phi  \\
& +  O\Big(  \int_0^{ |t|} ds 
  \Big [  \|    \phi   \|_{{\bf H}_{\langle x \rangle^{4 l}}^2(\mathbb{R}^3)^2}\Big(  \frac{1}{v ( 1 + |s|)^{2l -1}}  + 
\sum_{j = 1}^{2l}\frac{1}{(1 + |s|)^{2l - j} v^{\min(2, j)}} \Big) \notag \\  & \hspace{3cm}  + \| \phi \|_{{\bf H}^2(\mathbb{R}^3)^2} \frac{1}{v ( 1 + |s|)^{\zeta - 1}}  \Big ]  + \frac{v + |t|}{v^2} \| \phi \|_{{\bf H}_{\langle x \rangle^4}^{2}(\mathbb{R}^3)^2}   \Big).\noindent  
\end{align} 
\end{theorem}

\noindent{ \it Proof:}
By Duhamel's formula 
\begin{align}\label{e34}
  e^{i t H(\underline A)} \kappa  e^{-itH_{0}} F_{W}^{-1} e^{ix \cdot v \nu} \phi- 
\kappa F_{W}^{-1} e^{i x \cdot v \nu} & e^{i \int_{0}^{t} dr W_{\underline A}(r, \nu, \kappa)} \phi   \\
= &   \Big ( e^{i t H(\underline A)} \kappa  e^{-itH_{0}} F_{W}^{-1} e^{ix \cdot v \nu } e^{-i \int_{0}^{t}dr W_{\underline A}(r, \nu, \kappa)} 
  - \kappa F_{W}^{-1} e^{i x  \cdot v \nu } \Big ) e^{i \int_{0}^{t} dr W_{\underline A}(r, \nu, \kappa) } \phi   \notag \\ =
& \int_{0}^{t} ds \Big [  e^{is H(\underline A)} i( H(\underline A) \kappa -\kappa H_{0} ) 
e^{-is H_{0}} F_{W}^{-1}
e^{ix \cdot v \nu}  \phi (s, t) \notag \\ 
& \hspace{1cm} -  e^{is H(\underline A)} \kappa  e^{-is H_{0}} F_{W}^{-1} e^{i x \cdot v  \nu} i W_{\underline A} (s, \nu, \kappa)  \phi (s, t)\Big ]   
\notag,
\end{align}
where
\begin{align}
\phi(s,t) = e^{i \int_s^t dr W_{\underline A}(r, \nu, \kappa) }\phi.
\end{align} 
The desired result follows from Lemmata \ref{mero} and \ref{meroli} and Eq. \eqref{e34}, using that there is a constant $C$ such that 
\begin{equation}
\| \phi(s,t) \|_{{\bf H}^2_{\langle x\rangle^{4l} }(\mathbb{R}^3)^2} \leq C \| \phi \|_{{\bf H}^2_{\langle x\rangle^{4l} }(\mathbb{R}^3)^2}.  
\end{equation}
\bull
\begin{lemma}\label{nosta}
Let $\nu \in  \mathbb{S}^2 $.  
Suppose that $ \phi \in  {\bf H}_{\langle x \rangle^{2 \zeta }}^0(\mathbb{R}^3)^2$ is supported in $\Lambda_{\kappa, \nu}$. 
Let $ \underline A = (A_0, A)$,  with  $ A \in A_{\Phi}(B)$.
Then

\begin{align}\label{lolo}
\Big \| \big( e^{i \int_0^{\pm \infty} W_{\underline A}(r,  \nu, \kappa)}  - e^{i \int_0^{\pm t} W_{\underline A}(r, \nu, \kappa) } \big ) \phi \Big \|_{L^2(\mathbb{R}^3)^2} \leq \frac{C}{(1 + |t|)^{\zeta - 1}}
\| \phi \|_{ {\bf H}_{\langle x \rangle^{ 2\zeta  }}^0(\mathbb{R}^3)^2 },
\end{align}
for $t \geq 0 $.
\end{lemma}
\noindent{\it Proof:} We take the plus sign.  Let $\chi_t  $ be the characteristic function of the ball in $\mathbb{R}^3$
of radius $t$ and center $0$. Set $\phi_t = \chi_{t/2} \phi  $ and 
$\overline \phi_t  = 1 - \phi_t$. It is clear that 
\beq \label{clear}
\|   \overline \phi_{t}  \|_{L^2(\mathbb{R}^3)^2} \leq \frac{C}{(1 + |t|)^{\zeta }}
\| \phi \|_{ {\bf H}_{\langle x \rangle^{2 \zeta }}^0(\mathbb{R}^3)^2, }
\ene 
for some constant $C$. By Duhamel's formula and \eqref{clear}, 
\begin{align}\label{c2}
\Big \| \big( e^{i \int_0^\infty W_{\underline A}(r, \nu, \kappa)}  - e^{i \int_0^t W_{\underline A}(r, \nu, \kappa) } \big ) \phi \Big \|_{L^2(\mathbb{R}^3)^2}  \leq &
\Big \|  \int_t^{\infty} W_{\underline A}(r, \nu, \kappa) \phi_t \Big \|_{L^2(\mathbb{R}^3)^2} +
\Big \|  \int_t^{\infty} W_{\underline A}(r, \nu, \kappa)\overline  \phi_t \Big \|_{L^2(\mathbb{R}^3)^2}
\\  \notag  
\leq & \Big \|  \int_t^{\infty} W_{\underline A}(r,  \nu, \kappa) \phi_t \Big \|_{L^2(\mathbb{R}^3)^2} + 
\frac{C}{(1 + |t|)^{\zeta -1}}
\| \phi \|_{ {\bf H}_{\langle x \rangle^{\zeta }}^0(\mathbb{R}^3)^2}.
\end{align}
Next we notice, see \eqref{i33} and \eqref{e30}, that 
\beq \label{c3}
W_{\underline A}(r, \nu, \kappa) \phi_t =  e^{i r\beta \nu \cdot \mo    }W' e^{-i r \beta \nu \cdot \mo }\phi_t. 
\ene
As $  e^{-i r \beta \nu \cdot \mo }\phi_t $ is supported in the union of the balls or radius $t/2$ and centers $r \nu$ and $- r \nu$, the result follows from  \eqref{2.4} and \eqref{2.10b} (see also \eqref{c2} and \eqref{c3}).   

\bull

\begin{lemma}\label{L:i1}
Set $\nu \in  \mathbb{S}^2 $.
Suppose that $ \phi, \psi \in  {\bf H}^1(\mathbb{R}^3)^2$ are supported in $\Lambda_{\kappa, \nu} $. Let $ \underline A = (A_0, A)$,  with  $ A \in A_{\Phi}^{({\rm reg})}(B)$.
Then, for every $v \geq v_0$ (see Lemma \ref{pento})  

\begin{align}\label{i24}
 \Big | \langle \kappa F_{W}^{-1}e^{i x \cdot v \nu }\phi\:,  \kappa F_{W}^{-1}  
e^{i x \cdot v \nu} \psi \rangle_{{\cal H}(\underline A)}  -
\langle \kappa \phi, \: \kappa \psi \rangle_{L^{2}(\mathbb{R}^{n})^2} \Big | 
 \leq C  \frac{1}{v} \| \psi \|_{{\bf H}^{1}(\mathbb{R}^3)^2} \|\phi \|_{L^{2}(\mathbb{R}^{3})^2}. 
\end{align}

\end{lemma}

\noindent{\it Proof:}
Take 
\begin{align}
\phi = \begin{pmatrix} \phi_1, \phi_2 \end{pmatrix}, \hspace{1cm} \psi = \begin{pmatrix} \psi_1, \psi_2 \end{pmatrix}. 
\end{align}
It follows that
\begin{align}\label{t0}
 \langle \kappa F_{W}^{-1}e^{i x \cdot v \nu} \phi,\:    \kappa F_{W}^{-1}  
e^{i x \cdot v \nu} \psi \rangle_{{\cal H}(\underline A)} =  &  
 \langle B(\underline A) \kappa (\mo^{2}+ m^{2})^{-1/2} e^{i x \cdot v\nu} \phi_{1}, \:  
 B(\underline A)  \kappa 
(\mo^{2} + m^{2})^{-1/2}  
e^{i x \cdot v \nu } \psi_{1}  \rangle_{L^{2}(\mathbb{R}^{3})}  \\
& + 
\langle \kappa  \phi_{2}, \:    \kappa  \psi_{2} \rangle_{L^{2}(\mathbb{R}^{3})}, \notag
\end{align}
thus, is it enough to prove that 
\begin{align*}
& \Big | \langle B(\underline A)^{2} \kappa (\mo^{2}+ m^{2})^{-1/2}e^{i x \cdot v\nu } 
\phi_{1}, \:    \kappa (\mo^{2}+ m^{2})^{-1/2}  
e^{i x \cdot v\nu } \psi_{1} \rangle _{L^{2}(\mathbb{R}^{3})}  - \langle \kappa \phi_{1}, \: \kappa \psi_{1} \rangle_{L^{2}(\mathbb{R}^{3})} \Big | 
\leq C \frac{1}{v} \| \psi \|_{{\bf H}^{1}(\mathbb{R}^3)^2} \|\phi \|_{L^{2}(\mathbb{R}^{3})^2}. 
\end{align*} 
It is clear that 
\begin{align*}
 \langle B(\underline A)^{2} \kappa  (\mo^{2}+ m^{2})^{-1/2}e^{i x \cdot v \nu} \phi_{1}, \:    \kappa (\mo^{2}+ m^{2})^{-1/2} & 
e^{i x \cdot v \nu } \psi_{1} \rangle_{L^{2}(\mathbb{R}^{3})} \\ = & 
  \langle e^{-i x \cdot v\nu }( i\vartheta (\mo^{2}+ m^{2})^{-1/2})e^{i x \cdot v\nu} \phi_{1}, \:
 \kappa ((\mo + v\nu)^{2}+ m^{2})^{- 1/2} \psi_{1}\rangle_{L^{2}(\mathbb{R}^{3})} \\
& + \langle \kappa ((\mo + v\nu)^{2}+ m^{2})^{1/2}\phi_{1},\:
\kappa ((\mo + v\nu)^{2}+ m^{2})^{-1/2} \psi_{1} \rangle_{L^{2}(\mathbb{R}^{3})},
\end{align*}
where $ \vartheta  $ is defined in (\ref{vartheta}). \\
We notice that  $ \| e^{-i x \cdot v \nu }(\vartheta (\mo^{2}+ m^{2})^{-1/2})e^{i x \cdot v \nu } \|
_{{\cal L}(L^{2}(\mathbb{R}^{3}))} $ 
$ \leqq  C $ (independently of  $ v \nu$) and, furthermore, 
\begin{align} \label{t1}
\Big  \| \big ((\mo + v \nu )^{2}+ m^{2}\big )^{- 1/2} (\mo^{2}+ 1)^{-1 / 2} (\mo^{2}+ 1)^{1/2} \psi_{1} \Big \|_{L^{2}(\mathbb{R}^{3})} 
\leq C \frac{1}{v} \| \psi_{1} \|_{{\bf H}^{1}(\mathbb{R}^3)}.
\end{align} 
We obtain:  
\begin{align}\label{t2}
 \langle B(\underline A)^{2} \kappa (\mo^{2}+ m^{2})^{-1/2}e^{i x \cdot v \nu} \phi_{1}, \:    
\kappa (\mo^{2}+ m^{2})^{-1/2} & 
e^{i x \cdot v\nu } \psi_{1} \rangle_{L^{2}(\mathbb{R}^{3})}  \\
 = & \langle \kappa ((\mo + v\nu )^{2}+ m^{2})^{1/2}\phi_{1}, \:
\kappa ((\mo + v\nu)^{2}+ m^{2})^{-1/2} \psi_{1}\rangle_{L^{2}(\mathbb{R}^{3})} \notag \\
& + O\Big(\frac{1}{v}\Big) \| \psi_{1} \|_{{\bf H}^{1}(\mathbb{R}^3)} \|\phi_{1}\|_{L^{2}(\mathbb{R}^{3})}\notag
\end{align}  
Set $ \varsigma  := 1- \kappa^{2} $. The commutator
$ [ ((\mo + v\nu)^{2}+m^{2})^{1/2} , \varsigma  ] =  
 [ ((\mo + v\nu)^{2}+m^{2})^{1/2} , \kappa ^{2}
 ]  =  e^{-i x \cdot \nu v} [ ((\mo ^{2}+m^{2})^{1/2} , \varsigma  ] e^{i x \cdot \nu v} $ is bounded if and only if $  [ (\mo ^{2}+m^{2})^{1/2} , \varsigma  ]    $
 is bounded. The latter  
 is an  integral operator in momentum space. The kernel of this operator is
$$ (2 \pi )^{- 3} \widehat{ \varsigma  }(p- q) \Big( (p ^{2} + m^{2} )^{1/2} - 
(q^{2}+ m^{2})^{1/2} \Big ),  $$ 
 from which it is easy to see that 
\begin{align}\label{t3}
\| [ ((\mo + v\nu)^{2}+m^{2})^{1/2} , \kappa ^{2}  ] \|_{{\cal L}(L^{2}(\mathbb{R}^{3}))} =  \| [ (\mo ^{2}+m^{2})^{1/2} , \kappa ^{2}  ] \|_{{\cal L}(L^{2}(\mathbb{R}^{3}))} \leqq C 
\end{align}
where $C$ does not depend on $ v\nu $.\\
From (\ref{t1}), (\ref{t2}) and (\ref{t3}) we get 
\begin{align}\label{t4}
 \langle B(\underline A)^{2} \kappa (\mo^{2}+ m^{2})^{-1/2}e^{i x \cdot v\nu} \phi_{1}, \:   \kappa (\mo^{2}+ m^{2})^{-1/2}  
e^{i x \cdot v\nu} \psi_{1} \rangle_{L^{2}(\mathbb{R}^{3})}  
 = \langle \kappa \phi_{1}, \: \kappa \psi_{1} \rangle + O\Big(\frac{1}{v}\Big) \| \psi_{1} \|_{{\bf H}^{1}(\mathbb{R}^3)} \|\phi_{1}\|_{L^{2}(\mathbb{R}^{3})}. 
\end{align} 
Finally, (\ref{i24}) is obtained from (\ref{t0}) and (\ref{t4}).

\bull

\begin{theorem} \label{TP}
Set $\nu \in  \mathbb{S}^2 $ and $l \in \mathbb{N}$,  $l \geq \zeta/2, l \geq 2$.
Suppose that $ \phi, \psi \in  {\bf H}_{\langle x \rangle^{4 l}}^2(\mathbb{R}^3)^2$
are supported in $\Lambda_{\kappa, \nu} $. Let $ \underline A = (A_0, A)$,  with  $ A \in A_{\Phi}^{({\rm reg})}(B)$.
Then  
\begin{align} \label{TP1}
\langle e^{-i x \cdot v \nu } F_W  S (\underline{A})  F_W^{-1} e^{i x \cdot v\nu }\phi \; , 
\psi  \rangle_{L^2(\mathbb{R}^3)^2} =  &  \langle   e^{-i \int_{-\infty}^{\infty} W_{\underline A}(r, \nu, \kappa) dr } \phi \; , 
  \psi  \rangle_{L^{2}(\mathbb{R}^3)^2} 
 \\ & +  \|    \phi   \|_{{\bf H}_{\langle x \rangle^{4 l}}^2 (\mathbb{R}^3)^2}  
 \|    \psi   \|_{{\bf H}_{\langle x \rangle^{4 l}}^2 (\mathbb{R}^3)^2}
 \begin{cases} O\Big ( v^{1 - \zeta}  + \frac{1}{v} \Big ) & \text{if} 
 \: \zeta \ne 2 \\ \\
 O\Big ( \frac{ \ln(v)}{v}  \Big ) & \text{if} 
\: \zeta = 2.
 \end{cases} 
\end{align}

\end{theorem}
\noindent{\it Proof:}
We assume, without loss of generality, that $v \geq v_0$ (see Lemma \ref{pento}). 
Using Lemma \ref{meropp} and Theorem \ref{T:e0} we obtain
\begin{align} \label{TP1sta}
\langle e^{-i x \cdot v \nu } F_W  S (\underline{A})  F_W^{-1} e^{i x \cdot v\nu }\phi \; , 
\psi  \rangle_{L^2(\mathbb{R}^3)^2} = & \langle   W_{-}(\underline A)   F_W^{-1} e^{i x \cdot v\nu }\phi \; , 
 W_{+}(\underline{A}) F_W^{-1} e^{i x \cdot v\nu } \psi  \rangle_{\mathcal{H}(\underline A)} 
  \notag \\ \notag =  & \langle  e^{-it H(\underline A)} \kappa e^{itH_{0} }   F_W^{-1} e^{i x \cdot v\nu }\phi \; , 
 e^{it H(\underline A)} \kappa e^{-itH_{0} } F_W^{-1} e^{i x \cdot v\nu } \psi  \rangle_{\mathcal{H}(\underline A)} 
  \\ \notag & +   \|    \phi   \|_{{\bf H}_{\langle x \rangle^{4 l}}^2 (\mathbb{R}^3)^2}  
 \|    \psi   \|_{{\bf H}_{\langle x \rangle^{4 l}}^2 (\mathbb{R}^3)^2} 
O\Big ( \big ( \frac{1}{1 + |t|} \big )^{\zeta -1}  + \frac{1}{v^2} \Big )
\\ \notag =  &  \langle   \kappa F_{W}^{-1} e^{ix \cdot v \nu} e^{i \int_{0}^{-t} W_{\underline A}(r, \nu, \kappa) dr } \phi \; , 
 \kappa F_{W}^{-1} e^{ix \cdot v \nu} e^{i \int_{0}^{t} W_{\underline A}(r, \nu, \kappa) dr }  \psi  \rangle_{\mathcal{H}(\underline A)}
 \\ \notag & + \|    \phi   \|_{{\bf H}_{\langle x \rangle^{4 l}}^2 (\mathbb{R}^3)^2}  
 \|    \psi   \|_{{\bf H}_{\langle x \rangle^{4 l}}^2 (\mathbb{R}^3)^2} \Big\{ 
 O\Big ( \big ( \frac{1}{1 + |t|} \big )^{\zeta -1}  + \frac{1}{v^2} \Big ) 
 \\ \notag & \hspace{2cm}+  O\Big(  \int_0^{|t|} ds 
  \Big [    \Big(  \frac{1}{v ( 1 + |s|)^{2l-1}}  + 
\sum_{j = 1}^{2l}\frac{1}{(1 + |s|)^{2l - j} v^{\min(2, j)}} \Big) 
 \\ \notag  &
  \hspace{7.5cm}  +  \frac{1}{v ( 1 + |s|)^{\zeta - 1}}  \Big ]  + \frac{v + |t|}{v^2}    \Big) \Big \} 
\\  \notag  =  &  \langle   \kappa F_{W}^{-1} e^{ix \cdot v \nu} e^{i \int_{0}^{-t} W_{\underline A}(r, \nu, \kappa) dr } \phi \; , 
 \kappa F_{W}^{-1} e^{ix \cdot v \nu} e^{i \int_{0}^{t} W_{\underline A}(r, \nu, \kappa) dr }  \psi  \rangle_{\mathcal{H}(\underline A)} 
 \\ & +  \|    \phi   \|_{{\bf H}_{\langle x \rangle^{4 l}}^2 (\mathbb{R}^3)^2}  
 \|    \psi   \|_{{\bf H}_{\langle x \rangle^{4 l}}^2 (\mathbb{R}^3)^2}
 \begin{cases} O\Big ( v^{1 - \zeta}  + \frac{1}{v} \Big ) & \text{if} 
 \: \zeta \ne 2 \\ \\
 O\Big ( \frac{ \ln(v)}{v}  \Big ) & \text{if} 
\: \zeta = 2,
 \end{cases} 
\end{align}
where we select $ |t| =  v $, after integration, and we use that $l \geq 2$
. Eq. \eqref{TP1sta}, Lemma \ref{L:i1} and Lemma \ref{nosta} (taking $ |t| =  v $) imply that 
\begin{align} \label{casi}
\langle e^{-i x \cdot v \nu } F_W  S (\underline{A})  F_W^{-1} e^{i x \cdot v\nu }\phi \; , 
\psi  \rangle_{L^2(\mathbb{R}^3)^2}
    =  & 
  \langle    e^{i \int_{0}^{-t} W_{\underline A}(r, \nu, \kappa) dr } \phi \; , 
  e^{i \int_{0}^{t} W_{\underline A}(r, \nu, \kappa) dr }  \psi  \rangle_{L^2(\mathbb{R})^2} 
\\ & +  \|    \phi   \|_{{\bf H}_{\langle x \rangle^{4 l}}^2 (\mathbb{R}^3)}  
 \|    \psi   \|_{{\bf H}_{\langle x \rangle^{4 l}}^2 (\mathbb{R}^3)}
 \begin{cases} O\Big ( v^{1 - \zeta}  + \frac{1}{v} \Big ) & \text{if} 
 \: \zeta \ne 2 \\ \\
 O\Big ( \frac{ \ln(v)}{v}  \Big ) & \text{if} 
\: \zeta = 2 \end{cases} \notag
 \\ =  & 
  \langle    e^{i \int_{0}^{-\infty} W_{\underline A}(r, \nu, \kappa) dr } \phi \; , 
  e^{i \int_{0}^{\infty} W_{\underline A}(r, \nu, \kappa) dr }  \psi  \rangle_{L^2(\mathbb{R})^2} 
 \\ & +  \|    \phi   \|_{{\bf H}_{\langle x \rangle^{4 l}}^2 (\mathbb{R}^3)}  
 \|    \psi   \|_{{\bf H}_{\langle x \rangle^{4 l}}^2 (\mathbb{R}^3)}
 \begin{cases} O\Big ( v^{1 - \zeta}  + \frac{1}{v} \Big ) & \text{if} 
 \: \zeta \ne 2 \\ \\
 O\Big ( \frac{ \ln(v)}{v}  \Big ) & \text{if} 
\: \zeta = 2 .\end{cases}
\end{align}

\bull
\begin{remark}\label{rempw} { \rm
In the proof of Theorem \ref{TP} we use high-momenta asymptotic formulas for the wave operators, which are deduced in Eq. \eqref{TP1sta}. We actually prove:
\begin{align} \label{rempw1}
 \Big \| W_{-}(\underline A)   F_W^{-1} e^{i x \cdot v\nu }\phi  - 
   \kappa F_{W}^{-1} e^{ix \cdot v \nu} e^{i \int_{0}^{-\infty} W_{\underline A}(r, \nu, \kappa) dr } \phi \Big \|_{\mathcal{H}(\underline A)} = 
  \|    \phi   \|_{{\bf H}_{\langle x \rangle^{4 l}}^2 (\mathbb{R}^3)}  
 \begin{cases} O\Big ( v^{1 - \zeta}  + \frac{1}{v} \Big ) & \text{if} 
 \: \zeta \ne 2 \\ \\
 O\Big ( \frac{ \ln(v)}{v}  \Big ) & \text{if} 
\: \zeta = 2 \end{cases} , \\ \notag
\Big \|  W_{+}(\underline A)   F_W^{-1} e^{i x \cdot v\nu }\psi  - 
   \kappa F_{W}^{-1} e^{ix \cdot v \nu} e^{i \int_{0}^{\infty } W_{\underline A}(r, \nu, \kappa) dr } \psi  \Big \|_{\mathcal{H}(\underline A)}  = 
  \|    \psi   \|_{{\bf H}_{\langle x \rangle^{4 l}}^2 (\mathbb{R}^3)}  
 \begin{cases} O\Big ( v^{1 - \zeta}  + \frac{1}{v} \Big ) & \text{if} 
 \: \zeta \ne 2 \\ \\
 O\Big ( \frac{ \ln(v)}{v}  \Big ) & \text{if} 
\: \zeta = 2 \end{cases} .
\end{align}  }
\end{remark}

\subsubsection{High Momenta Limit for the Scattering Operator: General Magnetic Potentials (Proof of Theorem \ref{TPG})}\label{high-long}

In this section we prove our main results: We give a high-momenta expression for the scattering operator, with error bounds. This formula is the content of Theorem \ref{TPG}. Our formula is used in Sections
\ref{Sinv} and \ref{San} to reconstruct important information from the potentials and the magnetic field.

\paragraph{Proof of Theorem \ref{TPG}} 
 Eq. \eqref{change} implies that
$S(\underline A) = S(\underline{\tilde A})$ for  a regular magnetic potential $\tilde A \in \mathcal{A}^{{\rm reg}}_\Phi(B)$. Note that $ \tilde A$ always exists,  by Remark \ref{Rreg}, and furthermore, there is  $\lambda$  with $A - \tilde A  = \nabla \lambda$. In fact $\lambda $ is given by \eqref{claro}, from which we deduced in the lines below \eqref{2.12} that $\lambda_\infty$ is constant. This implies that
\begin{equation}\label{yaaaa}
 e^{-i \int_{-\infty}^{\infty} W_{\underline A}(r, \nu, \kappa) dr }  =  e^{-i \int_{-\infty}^{\infty} W_{\underline{ \tilde A}}(r, \nu, \kappa) dr } .
\end{equation}
It follows from \eqref{a5}-\eqref{diagu} and \eqref{i33} that
\begin{align}\label{cas}
W_{\underline{ A}}(t,  v\nu , \kappa ) = & e^{i t \beta(  \nu\cdot \mo )} \Big ( \kappa A_0(x) - \kappa   A \cdot \nu (x) \beta  \Big )e^{-i t \beta( \nu\cdot \mo )} \\ \notag = & Q^{-1}\begin{pmatrix}\kappa A_0(x + \nu t) 
-  \kappa  A \cdot \nu (x + \nu t) & 0 \\ 0 & \kappa A_0(x - \nu t) 
+  \kappa  A \cdot \nu (x - \nu t) \end{pmatrix} Q,
\end{align}  
and therefore
\begin{align} \label{in2}
   e^{-i \int_{-\infty}^{\infty} W_{\underline A }(r, v\nu, \kappa) dr } \phi  = &
 Q^{-1}
 \begin{pmatrix}
e^{i \int_{-\infty}^{\infty} dr \kappa (  A \cdot \nu - A_0) (x + r \nu ) }  & 0 
\\ 0 & e^{- i \int_{-\infty}^{\infty} dr \kappa  ( A \cdot \nu + A_0) (x - r \nu ) }
\end{pmatrix} Q \phi  \notag \\ 
 = &  
Q^{-1} \begin{pmatrix}
 e^{i \int_{-\infty}^{\infty} dr  (   A \cdot \nu - A_0) (x + r \nu ) }  & 0 
\\ 0 & e^{- i \int_{-\infty}^{\infty} dr   ( A \cdot \nu + A_0) (x + r \nu ) }
\end{pmatrix} Q \phi.
\end{align}     
The desired result follows from Theorem \ref{TP} and Eqs. \eqref{yaaaa} and \eqref{in2}.        

\bull

\subsection{Inverse-Scattering Reconstruction Method} \label{Sinv}

\subsubsection{ Proof of Theorem \ref{inverse-fields}}
Set $y \in \Lambda_{{\rm Rec}}$. Suppose that $P_y$ is a two-dimensional plane $P_y$ such that $y + P_y \subset  
\kappa^{-1}(\{ 1 \})^{\circ}$, for some function $\kappa \in 
C^\infty(\mathbb{R}^3) $ satisfying  \eqref{2.4} and the text above it (see Definition \ref{lrec} and Remark \ref{TRW}).  
 
Theorem \ref{TPG} implies that the scattering operator uniquely determines 

\begin{align} \label{in2aaa}
 \Big \langle
 \begin{pmatrix}
e^{i \int_{-\infty}^{\infty} dr  (   A\cdot \nu - A_0) (x + r \nu ) }  & 0 
\\ 0 & e^{- i \int_{-\infty}^{\infty} dr   ( A\cdot \nu + A_0) (x + r \nu ) }
\end{pmatrix} \phi,   \:
 \psi \Big \rangle_{L^{2}(\mathbb{R}^3)^2} .
\end{align} 
Here we suppose that $ \phi, \psi \in  
{\bf H}_{\langle x \rangle^{4 l}}^2(\mathbb{R}^3)^2$  ($l \in \mathbb{N}$,  $l \geq \zeta/2, l \geq 2$)   
 are supported in $ B(y; \delta) +  P_y$, for some small enough $\delta$ such that 
 $  B(y; \delta) +  P_y \subset  \kappa^{-1}(\{ 1\})^{\circ}  $. Selecting conveniently 
 $\phi$ and $\psi$ we obtain that the scattering matrix uniquely determines
\begin{align}\label{in3}
e^{-2 i \int_L  A  } = & \Big ( e^{i \int_{-\infty}^{\infty} dr  (   A\cdot \nu - A_0) (x + r \nu ) } \Big )^{-1} e^{- i \int_{-\infty}^{\infty} dr   ( A\cdot \nu + A_0) (x + r \nu ) }
\\ \notag 
e^{-i \int_L 2 A_0 } = &  e^{i \int_{-\infty}^{\infty} dr  (   A\cdot \nu - A_0) (x + r \nu ) }  e^{- i \int_{-\infty}^{\infty} dr   ( A\cdot \nu + A_0) (x + r \nu ) } ,  
\end{align}
for every line $L \subset y + P_y$. Here the integral in the first equality denotes an integral of the $1$-form 
$A$ and the one in the second denotes a scalar integral with respect to the Lebesgue measure in the line $L$. Denote by $R(A_0)(x, \nu) =  \int_{-\infty}^\infty A_0(x + \tau \nu)d\tau $.   Eq. \eqref{in3} implies that there exists an integer-valued function $n(x, \nu)$ such that  $2 R(x, \nu ) + 2 \pi n(x, \nu)$ can be recovered from the scattering operator, for every $x, \nu$ such that $x + \mathbb{R}\nu  \subset y + P_y$.  As $R$ is continuous, then $n $ can be taken to be constant. The decay properties of $A_0$ determine the value of $n$. Therefore we determine $R$ from the scattering operator. As $R$ describes a Radon transform, inverting this transform
we can uniquely reconstruct  $A_0$ in $y + P_y$. The full details of this procedure, as well as the reconstruction of the magnetic field, are carefully presented Theorem 6.3 in \cite{bw}. Using Theorem 6.3 in \cite{bw} we conclude that  $A_0(z)$ and $B(z)$ can be uniquely reconstructed from \eqref{in3}, for every $z \in y + P_y$. 
  
\bull

\subsection{The Aharonov-Bohm Effect} \label{San}
In this section we assume that $B =  0$ and $A_0 = 0$, i.e., that the electromagnetic field vanishes in $\Lambda$. The hypothesis $A_0 = 0$ is assumed for convenience, in the spirit of the Aharonov-Bohm effect. Nevertheless some results are also valid for $A_0 \ne 0$. In 
Section \ref{lala} we explain the corresponding results.  The results in this section are pretty much the same as the analogous achievements for the non-relativistic case in \cite{bw}. We follow the lines of \cite{bw}, Section 7, omitting repeated proofs. However, we must present again some notation, already introduced \cite{bw}, to help the reader to understand the statements of our results. In the case that $A_0 \ne 0$, notable differences between the relativistic and the non-relativistic cases hold true, see Section \ref{lala}.   
         
\subsubsection{Theorem \ref{th-7.1} and Applications} \label{T2.11}

\paragraph{Proof of Theorem \ref{th-7.1}} It follows from Theorem \ref{TPG}, \eqref{in2}  and the proof of Theorem    7.1 in \cite{bw}.
\bull 
\paragraph{Applications}  
  
\begin{remark} \label{rem-7.2}{\rm
Theorem \ref{th-7.1} implies that from the high-momenta limit (\ref{TP1I}) for $\hv $ and $\hw$ we can
reconstruct the fluxes
$$
\int_{\alpha}A
$$
for any closed curve $\alpha$ such that there is a surface (or chain) $\mathcal S$  in $\Lambda$ with
$\partial \mathcal S= \alpha -\gamma (x,y,\hv,\hw)$, because by Stokes' theorem

$$
 \int_{\alpha}A =\int_{\gamma (x,y,\hv,\hw)}A+ \int_{\mathcal S}B= \int_{\gamma (x,y,\hv,\hw)}A,
 $$
}
\end{remark}
\noindent since $B = 0$. 
We recall that (see \cite{gh}, page 47)
$
H_1(\Lambda;\ere)
$
represents the first group of singular homology of $\Lambda$ with
coefficients in $\mathbb{R}$. As the coefficients are in $\mathbb{R}$ it is, actually, a vector space.   We, furthermore, recall that  $
H^1_{\hbox{\rm de R}}(\Lambda)
$
represents the de Rahm cohomology group in $\Lambda$, see \cite{w}. 
\begin{remark} \label{rm-7.3}{\rm
As $\gamma(x,y,\hv,\hw)$ is a cycle, the homology class $[\gamma(x,y,\hv,\hw)]_{H_1(\Lambda;\ere)}$ is
well defined.

We denote by 
\beq
\hr:= \left\langle\left\{ [\gamma(x,y,\hv,\hw)]_{H_1(\Lambda;\ere)}: L(x, \hv)\cup L(x,\hw)\subset \Lambda
\right \}\right\rangle,
\label{7.3d}
\ene
where $\langle O \rangle $ denotes the vector space generated by $O$. 
 $\hr$ is a vector subspace of $H_1(\Lambda;\ere)$. Let us denote by $H^1_{ \hbox{\rm de  R, rec}}(\Lambda)$
the  vector subspace
of $H^1_{\hbox{\rm de  R}}(\Lambda)$ that is the dual to $\hr$, given by de Rham's theorem 
(Theorem 4.17, p. 154 of \cite{w}).
Then, for all $\Phi$ and all $A \in \0p2$, from the high-momenta limit (\ref{TP1I}) known for all $ \hv, \hw$
we reconstruct the projection of $A$ into $H^1_{\hbox{\rm de R, rec}}(\Lambda)$ modulo $2\pi$, as we now show.
Let
$$
\left\{ [\sigma_j]_{\hr} \right\}_{j=1}^m,
$$
be a basis of $\hr$, and let
$$
\left\{ [A_j]_{H^1_{\hbox{\rm de  R, rec}}(\Lambda)} \right\}_{j=1}^m,
$$
be the dual basis, i.e.,

$$
\int_{\sigma_j} A_k = \delta_{j,k}, j,k= 1,2, \cdots, m.
$$
Let us denote by $P_{\hbox{\rm rec}}$ the projector onto $H^1_{\hbox{\rm de  R, rec}}(\Lambda)$.
Hence, for any $A\in \0p2$
$$
 P_{\hbox{\rm rec}}\left[ A\right]_{H^1_{\hbox{\rm de  R}}(\Lambda)}=
 \sum_{j=1}^m \lambda_j  [A_j]_{H^1_{\hbox{\rm de  R, rec}}(\Lambda)},
$$
and, furthermore, as

$$
\lambda_j= \int_{\sigma_j} A,
$$
we reconstruct $\lambda_j, j=1,2,\cdots,m$ (modulo $ 2 \pi $ ) from the high-momenta limit (\ref{TP1I}) known for all $ \hv, \hw$.
}\end{remark}

\bull

\subsubsection{Theorem \ref{th-7.12I}} \label{T2.12}
In this section we prove Theorem \ref{th-7.12I}. This theorem is stated again in Theorem \ref{th-7.12}. We give additionally precise (explicit) definitions of the sets $\big \{  \Lambda_h  \big \}_{h \in \mathcal{I}}$, $\Lambda_{{\rm out}} $ and the numbers $\big \{  F_h  \big \}_{h \in \mathcal{I}}$, stated in Theorem \ref{th-7.12I}. We start by introducing some notations.

Below we give a definition of when a line $L(x,\hv)$ (see \eqref{line}) goes through holes of $K$. Take $r >0$ such that
$K \subset  B(0; r)$. Suppose that $  L(x,\hv) \subset \Lambda$, and
$L(x,\hv) \cap B(0; r) \neq \emptyset$. We denote by  $c(x,\hv)$ the curve consisting of the segment
$L(x,\hv) \cap \overline{B(0; r)}$ and  an arc on $ \partial\overline{ B(0; r)}$ that connects the points
$L(x,\hv) \cap \partial  \overline{ B(0; r)}$. We orient $c(x,\hv)$ in such a way that the segment of straight
line has the orientation of $\hv$. See Figure 2.

\begin{definition}[Definition 7.4 in \cite{bw}]\label{def-7.4} {\rm
A line  $L(x,\hv) \subset \Lambda$ goes through holes of $K$ if $L(x,\hv) \cap B(0; r) \neq \emptyset$
and $[c(x,\hv)]_{H_1(\Lambda;\ere)}\neq 0$. Otherwise we say that $L(x,\hv)$ does not go through holes of $K$. }
\end{definition}
%

\begin{definition}[Definition 7.5 in \cite{bw}] \label{def-7.5} {\rm
Two lines $L(x,\hv), L(y,\hw) \subset \Lambda$ that go through holes of $ K $  go through the same holes if $[c(x,\hv)]_{H_1(\Lambda;\ere)}=
\pm [c(y,\hw)]_{H_1(\Lambda;\ere)}$. Furthermore, we say that the lines go through the same holes in the same direction
if $[c(x,\hv)]_{H_1(\Lambda;\ere)}=
 [c(y,\hw)]_{H_1(\Lambda;\ere)}$. }
 \end{definition}

\begin{definition} \label{def-7.10} {\rm
For any $\hv \in \ese^2$ we denote by $ \Lambda_{\hv, \hbox{\rm out}}$ the set of points
$x \in \Lambda_{\hv}$ such that $L(x,\hv)$ does not go through holes of $K$. We call this set the region
without holes of $\Lambda_{\hv}$.  The holes of $\Lambda_{\hv}$ is the set $\Lambda_{\hv, \hbox{\rm in}}:=
\Lambda_{\hv} \setminus  \Lambda_{\hv, \hbox{\rm out}} $. }
\end{definition}

We define the following equivalence relation on $\Lambda_{\hv, \hbox{\rm in}}$. We say that
$ x R_{\hv} y$ if, and only if,
$L(x,\hv)$ and $L(y,\hv)$ go through the same holes and in the same direction. By $[x]$ we designate the
classes of equivalence under $R_{\hv}$.

We denote by $\left\{ \Lambda_{\hv, h} \right\}_{h \in \mathcal I}$ the partition of
$\Lambda_{\hv,\hbox{\rm in}}$ given by this equivalence relation. It is defined as follows.
$$
\mathcal I := \{ [x]  \}_{x\in \Lambda_{\hv, \hbox{\rm in}}}.
$$

Given $ h \in \mathcal I$ there is $x \in \Lambda_{\hv, \hbox{\rm in}}$ such that $h=[x]$. We denote,
$$
\Lambda_{\hv,h}:= \{y \in \Lambda_{\hv,\hbox{\rm in}}: y R_{\hv} x  \}.
$$
Then,
$$
\ds
\Lambda_{\hv, \hbox{\rm in}}= \bigcup_{h \in \mathcal I} \Lambda_{\hv,h},\,\,\,\,  \Lambda_{\hv,h_1}\cap
\Lambda_{\hv,h_2}= \emptyset, \,h_1 \neq h_2.
$$
We call $\Lambda_{\hv,h}$   the holes $h$ of $K$  in the direction of $\hv$.
Note that
\beq
\{\Lambda_{\hv,h}\}_{h \in \mathcal I} \cup \{\Lambda_{\v, \hbox{\rm out}}\}
\label{7.8}
\ene
is an open disjoint cover of $\Lambda_{\hv}$.

\begin{definition} \label{def-7.11} {\rm
For any $\Phi$, $ A \in \0p2$, $\hv \in \ese^2$,   and    $ h \in \mathcal I$ we define,
$$
F_h:= \int_{c(x,\hv)} A,
$$
 where $x$ is any point in $\Lambda_{\hv,h}$. Note that $F_h$ is independent  the    $x \in \Lambda_{\hv,h}$
that we choose. $F_h$ is the flux of the magnetic field over any surface (or chain) in $\ere^3$ whose boundary
is $c(x,\hv)$. We call $F_h$ the magnetic flux on the holes $h$ of $K$. }
\end{definition}

Let us take $\phi \in  {\bf H}_{\langle x \rangle^{4 l}}^2(\mathbb{R}^3)^2$ as in Theorem \ref{TPG}. We suppose, furthermore, that it is compactly supported. Then, since (\ref{7.8})
is a disjoint open cover of $\Lambda_{\hv}$,

\beq
\phi = \sum_{h\in \mathcal I} \varphi_h+ \varphi_{\hbox{\rm out}},
\label{7.9}
\ene
with $\varphi_h, \varphi_{\hbox{\rm out}} \in  {\bf H}_{\langle x \rangle^{4 l}}^2(\mathbb{R}^3)^2, \varphi_h$ has compact support
in $\Lambda_{\hv,h}, h\in \mathcal I$, and $\varphi_{\hbox{\rm out}}$ has compact support in
$\Lambda_{\hv,\hbox{\rm out}}$. The sum is finite because $\phi$ has compact support.
We denote, for $\v \in \mathbb{R}^3\setminus\{ 0 \}$ with $ \v/|\v| = \hv   $,
$$
\phi_{\v}:= e^{i\v\cdot x} \phi, \, \varphi_{h,\v}:= e^{i\v\cdot x}\varphi_h, \,
\varphi_{\hbox{\rm out}, \v}:= e^{i\v\cdot x} \varphi_{\hbox{\rm out}}.
$$
\begin{remark}\label{T} { \rm
We remark that for every flux $\Phi$   there exists a compactly supported $C^{\alpha}$-vector potential $\tilde A \in  \nb$. This holds true for the following reason: Take any $C^\infty$
gauge $A \in \nb$ (for example the Coulomb Gauge). Take any $C^\infty$ bounded domain  $\mathcal{D}$ containing $K$ such that $\ere^3 \setminus \mathcal{D}$ is simply connected. Set $x_0 \in \mathcal{D}^c$ and define 
\begin{align}\label{tiempo1}
\lambda(x) = \int_{C_{x_0, x}} A,
\end{align}
where $C_{x_0, x}$ is any $C^\infty$ curve connecting $x_0 $ with $x$ in $\mathcal{D}^c$. We extend $\lambda $ to a $C^{\alpha}$ function $ \bar \lambda $ defined in $\mathbb{R}^3$.  We take finally 
\beq \label{tiempo2}
\tilde A(x) = A(x) - \nabla \overline{\lambda}(x), \hspace{3cm} x \in \Lambda.   
\ene
  }

\end{remark}

\begin{theorem}\label{th-7.12} Set  $\phi, \psi \in  {\bf H}_{\langle x \rangle^{4 l}}^\alpha(\mathbb{R}^3)^2$ as in Theorem \ref{TPG}, with $\phi$ compactly supported. 
For every $A \in \nb$    

\begin{eqnarray}\label{fluxes}
\langle U S(\underline A) U^{-1}\, \phi_{\v}, \psi_{\v} \rangle = &  \Big \langle   \,\left( \sum_{h\in \mathcal I}\,
 \begin{pmatrix}  e^{i F_h}  & 0 \\ 0 &   e^{-i F_h} \end{pmatrix}  \varphi_{\v,h} + 
 \varphi_{\hbox{\rm out}, \v}\right) , \:   \psi_\v \Big \rangle\\ \notag & + O\left( \frac{1}{v}\right) \|    \phi   \|_{{\bf H}_{\langle x \rangle^{4 l}}^2 (\mathbb{R}^3)}  
 \| \psi \|_{{\bf H}_{\langle x \rangle^{4 l}}^2 (\mathbb{R}^3)} .
 \end{eqnarray}

\end{theorem}

\noindent {\it Proof:} 
Set $\tilde A  \in \nb$ be of class $C^2$, compactly supported, such that $  A = \tilde  A+  \nabla \lambda $, for some scalar function $\lambda$ (see \eqref{2.11} and Remark \ref{T}). Eq. \eqref{change} implies that $S(\underline A) = S(\underline{\tilde A})$, with $\underline A = (0, \tilde A)$. Then the desired result follows from Theorem \ref{TPG},  and the proof of Theorem 7.11 in \cite{bw}. Note that since $ \tilde A$ is of compact support the error term in 
Theorem \ref{TPG} is of order $O\left( \frac{1}{v}\right)$.
\begin{corollary} \label{cor-7.13}
Under the conditions of Theorem \ref{th-7.12},
the high-momenta limit \eqref{fluxes} of $S(\underline A)$ in  a single direction $\hv$  uniquely determines
the fluxes $ F_h, h\in \mathcal I$, modulo $2\pi$.
\end{corollary}
\noindent {\it Proof:} The corollary follows immediately from Theorem \ref{th-7.12}.

\subsubsection{The Case $A_0 \ne 0$ }\label{lala}
Taking $A_0 \ne 0$ does not change substantially our reasoning.
The results in Theorem \ref{th-7.1} and Remarks \ref{rem-7.2} and \ref{rm-7.3}
are deduced from the fact that the scattering operator uniquely determines   
\beq \label{A02}  
 e^{- i \int_{- \infty}^{\infty} A \cdot \nu (x + r \nu)}, 
\ene   
which in our case holds true only when $A_0 = 0$. However, if $A_0 \ne 0$, we can recover   
\beq \label{A01}
 e^{-i \int_{- \infty}^{\infty} 2A \cdot \nu (x + r \nu)},  
\ene 
see \eqref{in3}. Then if we substitute $A$ by $2 A$ in Theorem \ref{th-7.1} and Remarks \ref{rem-7.2} and \ref{rm-7.3} we obtain the same results, for $A_0 \ne 0$.  
Some care has to be taken while considering the function $\kappa$ as in \eqref{2.4}, it essentially amounts to substitute $K$ by $\kappa^{-1}(\{1\})^c$.
 We remark that the factor of $2$ signifies a notable difference from the non-relativistic case, in which this factor is not present, see \cite{bw}.  

Theorem \ref{th-7.12} gives a very simple formula for the high-momenta limit of the scattering operator in terms of magnetic fluxes. However, this simple formula is not anymore valid if $A_0 \ne 0$, because in this case  a factors of the from 
$e^{ \pm i \int_{- \infty}^{+ \infty } A_0(x + \tau \nu) d\tau}$ must be present, see \eqref{in2aaa}. This is also an important difference with respect to the non-relativistic case, in which the corresponding Theorem \ref{th-7.12}  is valid for $A_0 \ne 0$, see Theorem 7.11 in \cite{bw}.

\subsection{Some Technicalities: Stationary Phase Arguments}\label{tec}
In quantum mechanics the free evolution of particles follows the classical evolution up to some error. The probability of finding the particle in a certain region enclosing the classical trajectory can be estimated. The accuracy of finding the particle close to the place where the classical particle would be depends on the wave packet spreading.  These intuitive statements can be made precise by stationary phase arguments. Suppose for example that the free energy is given by the relativistic energy 
$ B_0 = (\mo^2 + m^2 )^{1/2} $  
 and the initial state is a wave function 
 $\phi \in \mathcal{S}(\mathbb{R}^3)$ whose Fourier transform is localized close to  $ p_0 $. The evolution of the particle at time $t$ is 
\beq\label{motiv1}
(e^{-i t B_0 } \phi)(x) = \frac{1}{(2\pi)^{3/2}} \int e^{i (x \cdot p - t (p^2 + m^2 )^{1/2})} \widehat \phi(p) dp.   
\ene 
Since  \eqref{motiv1} is an oscillatory integral, the bigger contribution is concentrated on the stationary points, i.e., the points on which the gradient in $p$ of the exponent vanishes:  
\beq \label{motiv2}
\nabla_p (x \cdot p - t (p^2 + m^2 )^{1/2}) = 0. 
\ene
If the support of $\hat \phi$ is contained in a small ball around    $p_0 $, this happens when 
\beq \label{motiv3}
x \simeq t \frac{p_0}{ (p_0^2 + m^2 )^{1/2}}, 
\ene
which is the description of the classical (relativistic) free trajectory with velocity
$ \frac{p_0}{ (p_0^2 + m^2 )^{1/2}} $.

 This motivates the following definition that associates the (relativistic) velocity to the momentum.
\begin{definition}\label{velocity}[Velocity Function] {\rm
We denote by ${\bf v} : \mathbb{R}^3  \mapsto \mathbb{R}^3$,
\beq \label{vel}
{\bf v}(p ) : = \frac{p}{(p^2 + m ^2)^{1/2}},
\ene
the function that associates to each momentum $p$ the corresponding velocity.  }
\end{definition}
If the particle is initially localized (at time $t=0$), to a good a approximation, in the ball    
$B \big (0;  |t| r_0/2\big )$, for some  $r_0 \in (0, \frac{1}{2} |x_0|)$ and $x_0 \in \mathbb{R}^3\setminus\{0\}$,  and the possible velocities are restricted (approximately)
to a ball $   B(x_0; r_0 ) $ then, at a time $t $, the particle is localized, to a good a approximation,  in the set $  B(0; |t|r_0/2 ) + t B(x_0; r_0) $ (the initial position plus the time times the velocity). This is the content of the next Lemma, which is based 
on Theorem XI.14 in \cite{rs3} (see also Lemma 2.1 in \cite{w1}).

\begin{lemma} \label{pento}
Take $x_0 \in \mathbb{R}^3 \setminus \{ 0 \} $, $r_0 \in (0, \frac{1}{2}\,|x_0|)$ and $f \in \mathcal{S}(\mathbb{R}^3)$  be such that $ {\bf v} \big( \operatorname{supp} ( f)\big ) \subset   B(x_0; r_0 )$. For every $l\in \mathbb{N}$ there is a constant $C_l$
such that 
\beq\label{masta}
\Big \|  \chi_{ \big ( B(0; |t|r_0/2 ) + t B(x_0; r_0)\big )^{\displaystyle{c}} } \: \cdot \:  e^{- i t B_0} f(\mo)
\chi_{B \big (0;  |t| r_0/2\big )}   \Big \| \leq C_{l} (1 + |t|)^{-l }. 
\ene
Moreover, let $\tau \in C_0^\infty(\mathbb{R}^3; [0, 1])$ be such that 
$\tau(x) = 1$ for $|x| \leq \frac{1}{2} $ and it vanishes for $ |x| \geq 1 $.  There exists 
$v_0 > 0$ and a constant $C_l$, for every $l \in \mathbb{N}$, such that
\beq\label{nonsta}
\Big \|  \chi_{ B(\nu t; |t|/2 )^{  \displaystyle{c}  } } \: \cdot \:   e^{- i t B_0} \tau \Big ( \frac{16(\mo - v \nu)}{v }\Big ) 
\chi_{B(0; |t|/8)}   \Big \| \leq C_{l} (1 + |t|)^{- l}, 
\ene
for every $\nu  \in \mathbb{S}^2$ and every $v \geq v_0$.
\end{lemma}
\noindent \emph{Proof:}
Let $\phi \in \mathcal{S}(\mathbb{R}^3)$. Using the Fourier transform we get 
\beq \label{ex1}
\Big [ e^{- i t B_0} f(\mo)
\chi_{B \big (0;  |t| r_0/2\big )}  \phi\Big ] (x) = \frac{1}{(2 \pi)^3} \int dy \chi_{B \big (0;  |t| r_0/2\big )}  \phi(y) \int dp e^{i[(x - y)\cdot p - t (p^2 + m^2)^{1/2} ]} f(p).    
\ene
We denote by 
\beq \label{ex2}
u_t(x, y) = \int dp e^{i[(x - y)\cdot p - t (p^2 + m^2)^{1/2} ]} f(p).  
\ene
The characteristic function in \eqref{masta} constrains the values of $x, y$ to satisfy
\beq \label{ex3}
y \in B \big (0;  |t| r_0/2\big ), \hspace{2cm} x \notin B(0; |t|r_0/2 ) + t B(x_0; r_0).   
\ene  
The Corollary to Theorem XI.14 in \cite{rs3} implies that for every 
$d \in \mathbb{N}$ there is a constant $c_d$ such that, for $x$ and $y$ satisfying \eqref{ex3},
\beq \label{ex4}
| u_t(x, y)| \leq c_d \frac{1}{( 1 + |t| + |x-y|)^d}.  
\ene
For $|x| \geq  |t| r_0 \geq 2 |y|$ we bound
\beq \label{ex5}
\frac{1}{( 1 + |t| + |x-y|)^d} \leq \frac{2^d}{( 1 + |t| + |x|)^d}.
\ene
and for $ |x| <  |t| r_0  $
\beq \label{ex6}
\frac{1}{( 1 + |t| + |x-y|)^d} \leq \frac{1}{( 1 + |t| )^d}.
\ene
Eqs. \eqref{ex1}-\eqref{ex6} and the Cauchy-Schwartz inequality imply that there is a constant $\tilde c_d$ such that
\beq \label{ex7}
\Big | \Big [ e^{- i t B_0} f(\mo)
\chi_{B \big (0;  |t| r_0/2\big )}  \phi\Big ] (x)\Big | \leq 
 \tilde c_d |t|^{3/2} \Big[\frac{1}{( 1 + |t| + |x|)^d} +  \frac{\chi_{B(0; |t|r_0)} }{( 1 + |t| )^d}   \Big ], 
\ene
 for $x$ satisfying \eqref{ex3}. A suitable election of $d$, the triangle inequality 
 and \eqref{ex7} imply \eqref{masta}. \\ 
Eq. \eqref{nonsta} follows from \eqref{masta} taking 
\beq \label{fv}
f_{v, \nu } \equiv f = \tau \Big ( \frac{16(\mo - v \nu)}{v }\Big ).  
\ene
The necessary hypotheses for \eqref{masta} are fulfilled for big $v$, taking $r_0 = 1/4$
and $ x_0 $ sufficiently close to $\nu$ to have 
\beq \label{ex8}
 \chi_{ \big ( B(0; |t|r_0/2 ) + t B(x_0; r_0)\big )^{ \displaystyle{c}} } 
\: \cdot \:  \chi_{ B(\nu t; |t|/2 )^{ \displaystyle{c}} }  = 
\chi_{ B(\nu t; |t|/2 )^{ \displaystyle{c}} } .
\ene
The fact that the constants are independent of $\nu$  and $ v$ follows from \eqref{ex1}-\eqref{ex8} changing the variable of integration $p$ in \eqref{ex1} by 
$  z = \frac{16(p - v \nu)}{v } $ and replacing $ x - y  $ by $ \frac{v}{16}(x - y) $
and $ t   $ by $ \frac{v}{16} t   $:
 \beq \label{ex2lol}
|u_t(x, y)| = \Big ( \frac{v}{16} \Big )^3 \Big |  \int dz \exp \Big [ i\big [ \big ((x - y)v/16\big ) \cdot  z - (tv/16) (z^2 + 32 z \cdot \nu  + 16^2 +  ( \frac{16}{v}  m)^2)^{1/2} \big ] \Big ] \tau(z) \Big | .  
\ene 
Then we apply the proof of the Corollary to Theorem XI.14 in \cite{rs3}. We point out that 
$\tau$ is independent of $ \nu  $ and  $v$.
Notice that we can assume that $ |t|\geq 1$ because 
the left hand sides of \eqref{masta} and \eqref{nonsta} are bounded (we take also $v_0 \geq 1$).

\bull

\begin{remark} \label{rpento} { \rm
The same conclusions of Lemma \ref{pento} hold true if we substitute  $ B_0 $ by $\nu \cdot \mo$ in \eqref{masta} and \eqref{nonsta}. Actually, although same proof can be applied, in this case the analysis is much simpler because
$e^{ - i t\nu \cdot \mo }$ is a translation operator in position:
\beq\label{mastamta}
\Big \|  \chi_{ \big ( B(0; |t|r_0/2 ) + t B(x_0; r_0)\big )^{\displaystyle{c}} } \: \cdot \:  e^{- i t \nu \cdot \mo } f(\mo)
\chi_{B \big (0;  |t| r_0/2\big )}   \Big \| \leq C_{l} (1 + |t|)^{-l }, 
\ene
\beq\label{nonstamta}
\Big \|  \chi_{ B(\nu t; |t|/2 )^{  \displaystyle{c}  } } \: \cdot \:   e^{- i t \nu \cdot \mo } \tau \Big ( \frac{16(\mo - v \nu)}{v }\Big ) 
\chi_{B(0; |t|/8)}   \Big \| \leq C_{l} (1 + |t|)^{- l}. 
\ene
We only prove \eqref{mastamta}:
Let $\phi \in \mathcal{S}(\mathbb{R}^3)$. Using the Fourier transform we get 
\beq \label{ex1mta}
\Big [ e^{- i t \nu \cdot \mo } f(\mo)
\chi_{B \big (0;  |t| r_0/2\big )}  \phi\Big ] (x) = \frac{1}{(2 \pi)^3} \int dy \chi_{B \big (0;  |t| r_0/2\big )}  \phi(y) \int dp e^{i[(x - y)\cdot p - t \nu \cdot \mo ]} f(p).    
\ene
We denote by 
\beq \label{ex2mta}
u_t(x, y) = \int dp e^{i[(x - y)\cdot p - t \nu \cdot \mo ]} f(p).  
\ene
The characteristic function in \eqref{mastamta} constrains the values of $x, y$ to satisfy
\beq \label{ex3mta}
y \in B \big (0;  |t| r_0/2\big ), \hspace{2cm} x \notin B(0; |t|r_0/2 ) + t B(x_0; r_0).   
\ene  
The Corollary to Theorem XI.14 in \cite{rs3} implies that for every 
$d \in \mathbb{N}$ there is a constant $c_d$ such that, for $x$ and $y$ satisfying \eqref{ex3},
\beq \label{ex4mta}
| u_t(x, y)| \leq c_d \frac{1}{( 1 + |t| + |x-y|)^d}.  
\ene
We finish the proof following \eqref{ex5}-\eqref{ex7}. 
}     
\end{remark}

\begin{corollary} \label{cocal}
Suppose that $V: \mathbb{R}^3 \to \mathbb{C}$ is such that 
\beq \label{upcot}
| V (x) | \leq C \frac{1}{(1 + |x|)^{\alpha_V}}
\ene
for some $  \alpha_V > 0 $.  Let $\phi \in \mathcal{S}(\mathbb{R}^3)$ be such that 
$f \hat \phi = \hat \phi$ for some $f$ satisfying the hypothesis of Lemma \ref{pento}.  
 Then there is a constant $C$ such that
\beq \label{upcot1}
\| V e^{i t B_0} \phi \| \leq C \Big ( \frac{1}{1 + |t|} \Big )^{\alpha_V }, \hspace{1cm } t \in \mathbb{R}. 
\ene
 
\end{corollary}

%

\noindent{\it Proof:}
The result is a direct consequence of Lemma \ref{pento} writing
$ V = \Big ( \chi_{ \big ( B(0; |t|r_0/2 ) + t B(x_0; r_0)\big )} + \chi_{ \big ( B(0; |t|r_0/2 ) + t B(x_0; r_0)\big )^c } \Big ) V   $ and 
$ \phi  = f(\mo )\phi = f(\mo ) \Big [ \chi_{B \big (0;  |t| r_0/2\big )} + \chi_{B \big (0;  |t| r_0/2\big )^c} \Big ] \phi $.

\bull

\subsubsection{Stationary Phase Arguments for High-Momenta} \label{high-stationary}
In this section we prove most of the technical results needed to derive the main achievements in this paper, which are stated and proved in Section \ref{main}. We estimate time evolution of relativistic particles, as explained at the beginning of Section \ref{tec}, with the particularity that the particles we consider are very energetic. Then they behave as classical particles moving in a ballistic way, up to an error bound. More precisely, we can 
substitute the relativistic evolution $  e^{- i t\big( (\mo + v\nu )^2  + m^2\big )^{1/2}}  $ by a translation operator $ e^{-it( v +\nu \cdot \mo) } $,  which represents a
classical free evolution with velocity $\nu$. Here $\| \nu \| = 1$; this is in agreement with the election of our units system in which the speed of light is set to $1$.

\noindent First we stress the following simple remark. We give the proof because it is used repeatedly
in this paper, although it is elementary.
\begin{remark}\label{obvious}  {\rm
There is a constant $C$ such that 
\beq \label{obv1}
\Big \|\big [ e^{- i t\big( (\mo + v\nu )^2  + m^2\big )^{1/2}}  - e^{-it( v +\nu \cdot \mo) }\big ] \phi \Big \| \leq 
C \frac{|t|}{ v} \| \phi \|_{{\bf H}^2(\mathbb{R}^3)},  
\ene  
and 
\beq \label{obv1mn}
\Bigg \|\Big | \frac{1}{\big ( (\mo + v\nu )^2  + m^2\big )^{1/2}}   \Big |    
\phi \Bigg \| \leq 
C \frac{1}{ v}\Big  \| \phi \|_{{\bf H}^1(\mathbb{R}^3)},   
\ene 
for every $v > 0$, $\nu \in \mathbb{S}^2$, $t \in \mathbb{R}$ and $\phi \in {\bf H}^2(\mathbb{R}^3)$. Furthermore, 
for every $\alpha, l \in \mathbb{N}$ there is a constant $C_l$ such that  
\beq \label{obv1que}
\Bigg \|\Big | \frac{\mo + v\nu}{\big ( (\mo + v\nu )^2  + m^2\big )^{1/2}}  - \nu \Big |^l    
\phi \Bigg \| \leq 
C_l \Big (\frac{1}{ v}\Big )^{\min(\alpha, l)} \| \phi \|_{{\bf H}^\alpha(\mathbb{R}^3)},  
\ene 
for every $v > 0$, $\nu \in \mathbb{S}^2$, and $\phi \in {\bf H}^\alpha(\mathbb{R}^3)$. }

\end{remark}
\noindent \emph{Proof:}
Using the fundamental theorem of calculus we find that
$|e^{ia} - a^{ib}| = | 1 - e^{ib - ia}  | = |\int_{0}^{b-a} e^{i\tau}d\tau| \leq |b-a| $; for every real numbers $a$, $b$. This implies  
Eq. \eqref{obv1}, since for $p \in \mathbb{R}^3$
\beq \label{obv2}
\Big \| \big [ e^{- i t\big( (p + v\nu )^2  + m^2\big )^{1/2}}  - e^{-it ( v +\nu \cdot p )}\big ] \langle p \rangle^{-2} \Big \|
\leq C \frac{|t|}{ v},
\ene that is a consequence of the next calculation: 
\beq \label{obv3}
 \Big |\big ((p + v\nu )^2  + m^2\big )^{1/2}  -  ( v +\nu \cdot p )\Big | \langle  p \rangle^{-2 }=  \Bigg |\frac{ p^2 + m^2 - (\nu \cdot p)^2 }{ \big( (p + v\nu )^2  + m^2\big )^{1/2}  +  ( v +\nu \cdot p )  }\Bigg |\langle p \rangle^{-2 }.
\ene
We estimate \eqref{obv3} separately for $|p| < v/2$ and $|p| \geq  v/2$. For $|p| \geq v /2$ we use the left hand side 
of \eqref{obv3} taking advantage of $ \langle  p \rangle^{-2 }  $. For  $|p| < v/2$ we estimate the right hand side, taking into account that the denominator is bounded from below by $ v/2$ and that $\Big | p^2 + m^2 - (\nu \cdot p)^2 \Big |\langle p \rangle^{-2 }$ is uniformly bounded.   \\
Now we prove \eqref{obv1que}. We proceed as before taking $p \in \mathbb{R}^3$ instead of $\mo$. We write $\phi =\langle  \mo \rangle^{-\alpha } \langle  \mo \rangle^{\alpha } \phi  $. For $|p| \geq v /2$ we take advantage of $ \langle  p \rangle^{-\alpha }  $ as before. For  $|p| < v /2$ we use  
\begin{align} \label{mta1}
\Bigg | \frac{p + v\nu}{\big ( (p + v\nu )^2  + m^2\big )^{1/2}}  - \nu \Bigg | \leq &
\Bigg | \frac{p }{\big ( (p + v\nu )^2  + m^2\big )^{1/2}}   \Bigg | + 
\Bigg | \frac{    \Big ( \big ( (p + v\nu )^2  + m^2\big )^{1/2} -  ( v +\nu \cdot p )\Big ) \nu  }{\big ( (p + v\nu )^2  + m^2\big )^{1/2}   }   \Bigg |  
\\ \notag & + \Bigg | \frac{ (\nu \cdot p)  \nu  }{\big ( (\mo + v\nu )^2  + m^2\big )^{1/2}   }   \Bigg |  
\end{align} 
and \eqref{obv3}, without the factor $ \langle p \rangle^{-2}  $. Notice that in this case 

$$\Bigg |\frac{ p^2 + m^2 - (\nu \cdot p)^2 }{ \big( (p + v\nu )^2  + m^2\big )^{1/2} 
 +  ( v +\nu \cdot p )  }\Bigg | \leq  |p| \Bigg | \frac{ v }{ \big( (p + v\nu )^2  + m^2\big )^{1/2} 
 +  ( v +\nu \cdot p )  }\Bigg | + m \leq C (|p| + 1),$$  
for some constant $C$. Using similar techniques we prove Eq. \eqref{obv1mn}.

\bull

\begin{lemma} \label{pen1}
Let $\tau \in C_0^\infty(\mathbb{R}^3; [0, 1])$ be such that 
$\tau(x) = 1$ for $|x| \leq \frac{1}{2} $ and it vanishes for $ |x| \geq 1 $. Let $h : \mathbb{R}^3 \mapsto \mathbb{R}$ satisfy 
\beq\label{pen20}
|h(x)| \leq C \Big (\frac{1}{1 + |x|}\Big )^{\zeta},
\ene
for some $\zeta > 1$ and some constant $C$. Take $f \in C^\infty(\mathbb{R}^n) $ bounded with all derivatives bounded. For every $\alpha, l \in \mathbb{N}$ with $\alpha \geq 2$ there is a constant $C_{l}$ such that
\begin{align} \label{he1}
\Big \| h(x) f(\mo + v \nu) \Big[ e^{ - it \big( (\mo + v \nu)^2 + m^2\big )^{1/2}} - e^{ - it (v + \nu \cdot \mo)}\Big ] \tau\Big( \frac{16 \mo}{v} \Big)\phi \Big \| \leq 
C_{l}  \Big [  \|    \phi   \|_{{\bf H}_{\langle x \rangle^{4 l}}^\alpha(\mathbb{R}^3)}\Big(  \frac{1}{v ( 1 + |t|)^{2l -1}} & + 
\sum_{j = 1}^{2l}\frac{1}{(1 + |t|)^{2l - j} v^{\min(\alpha, j)}} \Big) \notag \\  &  + \| \phi \|_{{\bf H}^2(\mathbb{R}^3)} \frac{1}{v ( 1 + |t|)^{\zeta - 1}}  \Big ], 
\end{align}
for every $v > v_0$ (see Lemma \ref{pento}), $\nu \in \mathbb{S}^2$, $t \in \mathbb{R}$, and every $\phi \in {\bf H}_{\langle x \rangle^{4 l}}^\alpha(\mathbb{R}^3) $.  
\end{lemma}

\noindent \emph{Proof:} We take $t > 0$ (without loss of generality).  We use the shorthand notation 
\beq \label{he2}
\psi  = h(x) f(\mo + v \nu) \Big[ e^{- it \big( (\mo + v \nu)^2 + m^2\big )^{1/2}} - e^{- it (v + \nu \cdot \mo)}\Big ] \tau\Big( \frac{16 \mo}{v} \Big)\phi
\ene
and write 
\beq \label{he3}
\psi = \psi^{(1)} + \psi^{(2)}
\ene
with
\beq \label{he4}
\psi^{(1)} = \chi_{B(\nu t; |t|/2 )} \psi,  \hspace{1cm} \psi^{(2)} =  \chi_{B(\nu t; |t|/2 )^{ \displaystyle{c}} } \: \cdot \:  \psi.
\ene
We estimate first $\psi^{(1)}$:
\begin{align} \label{he5}
\| \psi^{(1)}\| \leq & \| \chi_{B(\nu t; |t|/2 )}   h(x) f(\mo + v \nu)  \|\cdot \big \|
e^{- it \big( (\mo + v \nu)^2 + m^2\big )^{1/2}} - e^{- it (v + \nu \cdot \mo)}\Big ] \tau\Big( \frac{16 \mo}{v} \Big)\phi \big \|
\\ \leq &  \notag C \frac{1}{v} \Big ( \frac{1}{1 + |t|}\Big )^{\zeta - 1} \| \phi \|_{{\bf H}^2(\mathbb{R}^3)},
\end{align}
where use Remark \ref{obvious} and \eqref{pen20}.  \\
Now we estimate $\psi^{(2)}$. We have that 
\begin{align} \label{he6}
\| \psi^{(2)} \| \leq C & \Big [ \Big \|  \chi_{B(\nu t; |t|/2 )^{   \displaystyle{c}   }} \: \cdot \:  e^{- it \big( (\mo + v \nu)^2 + m^2\big )^{1/2}} \tau
\Big ( \frac{16 \mo }{v}\Big )\frac{\chi_{B(0; |t|/8 )}}{\langle x \rangle ^{2l}}   \Big \|
+ \Big \| \frac{\chi_{B(0; |t|/8 )^{   \displaystyle{c}}   }}{\langle x \rangle ^{2l}}   \Big \|   \Big ]
\\ \notag & \cdot \Big \| \Big[ \langle x \rangle^{2l} f(\mo + v \nu) 
\Big[ e^{ it \big [ \big( (\mo + v \nu)^2 + m^2\big )^{1/2}  -  (v + \nu \cdot \mo)\big ] } - 1  \Big  ] \phi \Big \| \\ \notag
\leq C_l & \Big (  \frac{1}{1 + |t|}\Big )^{2l}  \Big \| \Big[ \langle x \rangle^{2l} f(\mo + v \nu) 
\Big[ e^{- it \big [ \big( (\mo + v \nu)^2 + m^2\big )^{1/2}  -  (v + \nu \cdot \mo)\big ] } - 1  \Big  ] \phi \Big \|, 
\end{align}
where we use Lemma \ref{pento} and \eqref{usefullEq}. To estimate the remaining part in \eqref{he6} we notice that 
the operator $x$ is a differential operator in the Fourier transform representation (or momentum space): 
\beq  \label{nablap}
x =  i \nabla_{\mo}.
\ene
Taking the commutator of $\langle x \rangle^{2l}$ with functions of $\mo$ produces derivatives with respect to $\mo$. Taking into consideration that all derivatives of $f$ are bounded and using \eqref{obv1que} (and similar estimates) we get
\begin{align} \label{he7}
\Big \| \Big[ \langle x \rangle^{2l} f(\mo + v \nu) 
\Big[ e^{ it \big [ \big( (\mo + v \nu)^2 + m^2\big )^{1/2}  -  (v + \nu \cdot \mo)\big ] } - 1  \Big  ] \phi \Big \| \leq & C \Big \|  \Big[ \langle x \rangle^{2l}\: ,\: f(\mo + v \nu) 
\Big( e^{ it \big [ \big( (\mo + v \nu)^2 + m^2\big )^{1/2}  -  (v + \nu \cdot \mo)\big ] } - 1 \Big )  \Big  ] \phi \Big \| \notag
\\ \notag  & + C \Big \|   f(\mo + v \nu) 
\Big[ e^{ it \big [ \big( (\mo + v \nu)^2 + m^2\big )^{1/2}  -  (v + \nu \cdot \mo)\big ] } - 1  \Big  ]\langle x \rangle^{2l} \phi \Big \|. 
\\ \leq & C \sum_{j = 1}^{2l} \Big ( \frac{1 + |t|}{v}\Big )^{\min(\alpha, j)} \|    
\phi  \|_{{\bf H}_{\langle x \rangle^{4l}}^\alpha(\mathbb{R}^3)}
+ C \frac{1 + |t|}{v}\|   \phi  \|_{{\bf H}_{\langle x \rangle^{4l}}^\alpha(\mathbb{R}^3)},
\end{align}
Where we use the fact that for every $  g\in {\bf H}_{\langle x \rangle^{4l}}^\alpha(\mathbb{R}^3)$ and every 
multi-index $\underline l = (l_1, l_2, l_3) \in (\mathbb{N}\cup \{0\})^3 $, with $l_1 + l_2 + l_3 \leq 2l$,
\beq \label{noquier}
\| \langle \mo \rangle^{\alpha} x_1^{l_1} x_2^{l_2} x_3^{l_3} g \| \leq C \|  x_1^{l_1} x_2^{l_2} x_3^{l_3} g \|_{{\bf H}^{\alpha}(\mathbb{R}^3)} \leq C \|    
g  \|_{{\bf H}_{\langle x \rangle^{4l}}^\alpha(\mathbb{R}^3)},
\ene
for some constant $C$. Eq. \eqref{he1} follows from \eqref{he2},  \eqref{he3}, \eqref{he5}, \eqref{he6} and \eqref{he7}.

\bull

\begin{lemma} \label{pen1o}
Let $\tau \in C_0^\infty(\mathbb{R}^3; [0, 1])$ be such that 
$\tau(x) = 1$ for $|x| \leq \frac{1}{2} $ and it vanishes for $ |x| \geq 1 $. Let $h : \mathbb{R}^3 \mapsto \mathbb{R}$ satisfy 
\beq\label{pen2}
|h(x)| \leq C \Big  (\frac{1}{1 + |x|}\Big )^{\zeta},
\ene
for some $\zeta > 1$ and some constant $C$. Take $f \in C^\infty(\mathbb{R}^n) $ bounded with all derivatives bounded. For every $ l \in \mathbb{N}$ there is a constant $C_{l}$ such that
\begin{align} \label{he1o}
\Big \| h(x) f(\mo + v \nu) \Big[ e^{- it \big( (\mo + v \nu)^2 + m^2\big )^{1/2}} \Big ] \tau\Big( \frac{16 \mo}{v} \Big)\phi \Big \| \leq 
C_{l}  \Big [  \|    \phi   \|_{{\bf H}_{\langle x \rangle^{4 l}}^0(\mathbb{R}^3)}\Big(  \frac{1}{ ( 1 + |t|)^{2l}} &  \Big) 
 + \| \phi \|_{{\bf H}^0(\mathbb{R}^3)} \frac{1}{ ( 1 + |t|)^{\zeta }}  \Big ], 
\end{align}
for every $v > v_0$ (see Lemma \ref{pento}), $\nu \in \mathbb{S}^2$, $t \in \mathbb{R}$, and every $\phi \in {\bf H}_{\langle x \rangle^{4 l}}^0(\mathbb{R}^3) $.  
\end{lemma}

\noindent \emph{Proof:} We take $t > 0$ (without loss of generality). We follow the procedure of the proof of Lemma \ref{pen1}.   We use the shorthand notation 
\beq \label{he2o}
\psi  = h(x)( f(\mo + v \nu) \Big[ e^{- it \big( (\mo + v \nu)^2 + m^2\big )^{1/2}} \Big ] \tau\Big( \frac{16 \mo}{v} \Big)\phi
\ene
and write 
\beq \label{he3o}
\psi = \psi^{(1)} + \psi^{(2)}
\ene
with
\beq \label{he4o}
\psi^{(1)} = \chi_{B(\nu t; |t|/2 )} \psi,  \hspace{1cm} \psi^{(2)} =  \chi_{B(\nu t; |t|/2 )^c} \psi.
\ene
We estimate first $\psi^{(1)}$:
\begin{align} \label{he5o}
\| \psi^{(1)}\| \leq &  \| \chi_{B(\nu t; |t|/2 )}   h(x) f(\mo + v \nu)  \|\cdot \big \|
e^{- it \big( (\mo + v \nu)^2 + m^2\big )^{1/2}}  \tau\Big( \frac{16 \mo}{v} \Big)\phi \big \|
\\ \leq &  \notag  C \Big ( \frac{1}{1 + |t|}\Big )^{\zeta } \| \phi \|_{{\bf H}^0(\mathbb{R}^3)},
\end{align}
where use \eqref{pen2}.  \\
Now we estimate $\psi^{(2)}$. We have that 
\begin{align} \label{he6o}
\| \psi^{(2)} \| \leq C & \Big [ \Big \|  \chi_{B(\nu t; |t|/2 )^c} e^{- it \big( (\mo + v \nu)^2 + m^2\big )^{1/2}} \tau
\Big ( \frac{16 \mo }{v}\Big )\frac{\chi_{B(0; |t|/8 )}}{\langle x \rangle ^{2l}}   \Big \|
+ \Big \| \frac{\chi_{B(0; |t|/8 )^c}}{\langle x \rangle ^{2l}}   \Big \|   \Big ]
\cdot 
\Big \| \Big[ \langle x \rangle^{2l} f(\mo + v \nu) 
 \phi \Big \| \\ \notag
\leq C_l & \Big (  \frac{1}{1 + |t|}\Big )^{2l}  \Big \| \Big[ \langle x \rangle^{2l} f(\mo + v \nu)  \phi \Big \|, 
\end{align}
where we use Lemma \ref{pento} and \eqref{usefullEq}. To estimate the remaining part in \eqref{he6} we use that $x  = i \nabla_{\mo}$.  
Taking the commutator of $\langle x \rangle^{2l}$ with functions of $\mo$ produces derivatives with respect to $\mo$ and as all derivatives of $f$ are bounded we get 
\begin{align} \label{he7o}
\Big \| \Big[ \langle x \rangle^{2l} f(\mo + v \nu)  \phi \Big \| \leq & \Big \|  \Big[ \langle x \rangle^{2l}\: ,\: f(\mo + v \nu)   \Big  ] \phi \Big \| 
 + \Big \|   f(\mo + v \nu) \langle x \rangle^{2l} \phi \Big \| 
\leq 
C \|   \phi  \|_{{\bf H}_{\langle x \rangle^{4l}}^0(\mathbb{R}^3)}.
\end{align}
Eq. \eqref{he1o} follows from \eqref{he2o},  \eqref{he3o}, \eqref{he5o}, \eqref{he6o} and \eqref{he7o}.

\bull

\begin{lemma} \label{pent1}
Let $\tau \in C_0^\infty(\mathbb{R}^3; [0, 1])$ be such that 
$\tau(x) = 1$ for $|x| \leq \frac{1}{2} $ and it vanishes for $ |x| \geq 1 $. 
Let
 $ h  \in C^{2}(\mathbb{R}^3)$ satisfy 
\beq\label{pent2}
\Big |\Big ( \frac{\partial}{\partial x_1}\Big )^{l_1}\Big ( \frac{\partial}{\partial x_2}\Big )^{l_2}\Big ( \frac{\partial}{\partial x_3}\Big )^{l_3} h(x)\Big | \leq C \Big (\frac{1}{1 + |x|}\Big )^{\zeta},
\ene
for some $\zeta > 1$, a constant $C$ and every $l_1, l_2, l_3 \in \mathbb{N}\cup \{  0\}$ with $l_1 + l_2 + l_3 \leq 2$. Take $f \in C^\infty(\mathbb{R}^n) $ bounded with all derivatives bounded. For every $l \in \mathbb{N}$, there is a constant $C_l$ such that
\begin{align} \label{het1}
\Big \|\Big[ e^{- it \big( (\mo + v \nu)^2 + m^2\big )^{1/2}} -   e^{- it (v + \nu \cdot \mo)}\Big ] &  h(x) f(\mo + v \nu)e^{- it (v + \nu \cdot \mo)} \tau\Big( \frac{16 \mo}{v} \Big)\phi \Big \|    \leq C \frac{1}{v} \Big (\frac{1}{1 + |t|}\Big)^{\zeta - 1} \|  \phi \|_{{\bf H}^{2}(\mathbb{R}^3)} \\ & \hspace{6cm} \notag + C_l 
\frac{1}{v} \Big (\frac{1}{1 + |t|}\Big)^{2l-1} \|  \phi \|_{{\bf H}^{2}_{\langle x \rangle^{4l} }(\mathbb{R}^3)},
\end{align}
for every $v > v_0$ (see Lemma \ref{pento}), $\nu \in \mathbb{S}^2$, $t \in \mathbb{R}$, and every $\phi \in {\bf H}^{2}_{\langle x \rangle^{4l} }(\mathbb{R}^3) $.  
\end{lemma}

\noindent \emph{Proof:} We take $t > 0$. 
Let $\theta \in C_0^\infty(\mathbb{R}^3; [0, 1])$ be equal $1$ in $B(\nu; 1/2)$ and zero in $B(\nu; 2/3)^c$. Define 
\beq
\theta_t(x) : = \theta( x/t).  
\ene
We use the shorthand 
\begin{align} \label{het2}
\psi^{(1)} & = \Big[ e^{- it \big( (\mo + v \nu)^2 + m^2\big )^{1/2}} -   e^{- it (v + \nu \cdot \mo)}\Big ] \theta_t(x)  h(x) f(\mo + v \nu)e^{- it (v + \nu \cdot \mo)} \tau\Big( \frac{16 \mo}{v} \Big)\phi , 
\\ \notag 
\psi^{(2)} & = \Big[ e^{- it \big( (\mo + v \nu)^2 + m^2\big )^{1/2}} -   e^{- it (v + \nu \cdot \mo)}\Big ](1 - \theta_t(x))   h(x) f(\mo + v \nu)e^{- it (v + \nu \cdot \mo)} \tau\Big( \frac{16 \mo}{v} \Big)\phi. 
\end{align}
We analyze first $\psi^{(1)}$. He have that
\begin{align}\label{het3}
\| \psi^{(1)} \| \leq &\Big \|  \Big[ e^{- it \big( (\mo + v \nu)^2 + m^2\big )^{1/2}} -   e^{- it (v + \nu \cdot \mo)}\Big ]  \langle \mo \rangle^{- 2}   \Big \|\cdot \Big \|   \langle \mo \rangle^{ 2}\theta_t(x)  h(x) f(\mo + v \nu)e^{- it (v + \nu \cdot \mo)} \tau\Big( \frac{16 \mo}{v} \Big)\phi  \Big \|
\\  \leq & C \frac{|t|}{v} \Big (\frac{1}{1 + |t|}\Big)^{\zeta} \|  \phi \|_{{\bf H}^{2}(\mathbb{R}^3)},
\end{align}
where we use \eqref{obv2} and \eqref{pent2}. \\
Now we estimate $\psi^{(2)}$ using Remark \ref{rpento} and \eqref{pent2}:  
\begin{align}\label{het4}
\| \psi^{(2)} \| \leq &\Big \|  \Big[ e^{- it \big( (\mo + v \nu)^2 + m^2\big )^{1/2}} -   e^{- it (v + \nu \cdot \mo)}\Big ]  \langle \mo \rangle^{- 2}   \Big \|\cdot \Big \|   \langle \mo \rangle^{ 2}    (1 - \theta_t(x))  h(x) f(\mo + v \nu)e^{- it (v + \nu \cdot \mo)} \tau\Big( \frac{16 \mo}{v} \Big)\phi  \Big \|
\\  \leq & 
C_l 
\frac{|t|}{v} \Big (\frac{1}{1 + |t|}\Big)^{2l} \|  \phi \|_{{\bf H}^{2}_{\langle x \rangle^{4l} }(\mathbb{R}^3)},
\end{align}
for every $l \in \mathbb{N}$.

\bull

\begin{lemma}\label{estep}
Let $h : \mathbb{R}^3 \mapsto \mathbb{R}$ and $b : \mathbb{R}^3 \mapsto \mathbb{R}^3$ be such that 
\beq
|h(x)| \leq C \Big ( \frac{1}{1 + |x|}\Big )^{\zeta}, \hspace{1cm}\| b(x)\| \leq  C \Big ( \frac{1}{1 + |x|}\Big )^{\zeta},
\ene
for some constant $ C$, and every $x \in \mathbb{R}^3$. 
It follows that the exist a constant $C$ such that
\begin{align}\label{cansa1}
\Bigg \| h(x) \frac{1}{ \big ((\mo + v\nu)^2 + m^2 \big )^{1/2} } e^{- i t( v + \nu \cdot \mo)}\phi \Bigg \| \leq
C\Big [ \frac{1}{v (1+ |t|)^{\zeta}} \| \phi \|_{{\bf H}_{\langle x \rangle^{4 l}}^1(\mathbb{R}^3)} + 
  \frac{1}{v^{\alpha}} \| \phi \|_{{\bf H}^\alpha(\mathbb{R}^3)} \Big ]
\end{align}
and
\begin{align}\label{cansa2}
\Bigg \| b(x) \cdot \Big[ \frac{\mo + v \nu }{ \big ((\mo + v\nu)^2 + m^2 \big )^{1/2} } - \nu \Big ] e^{- i t( v + \nu \cdot \mo)}\phi \Bigg \| \leq C\Big [ \frac{1}{v (1+ |t|)^{\zeta}} \| \phi \|_{{\bf H}_{\langle x \rangle^{4 l}}^1(\mathbb{R}^3)} + \frac{1}{v^{\alpha}}
\| \phi \|_{{\bf H}^\alpha(\mathbb{R}^3)} \Big ],
\end{align}
for every $v > v_0$ (see Lemma \ref{pento}), $\nu \in \mathbb{S}^2$, $t \in \mathbb{R}$, every natural number $l \geq \zeta/2$, and every $\phi \in {\bf H}_{\langle x \rangle^{4 l}}^\alpha(\mathbb{R}^3) $, with $\alpha \in  \{1, 2,\cdots \} $.

\end{lemma}
\emph{Proof:} We take, without loss of generality, $t  > 0$.  
Let $\tau \in C_0^\infty(\mathbb{R}^3; [0, 1])$ be such that 
$\tau(x) = 1$ for $|x| \leq \frac{1}{2} $ and it vanishes for $ |x| \geq 1 $.
Notice that 
\beq \label{cansa3}
\| (1 - \tau(16 \mo/v)) \phi \| \leq C \frac{1}{v^\alpha} \| \phi \|_{{\bf H}^{\alpha}(\mathbb{R}^3)}. 
\ene 
We prove \eqref{cansa1}. The proof of \eqref{cansa2} is similar, using \eqref{obv1que}. Define 
\beq
\label{cansa4}
\psi =  h(x) \frac{1}{ \big ((\mo + v\nu)^2 + m^2 \big )^{1/2} } e^{- i t (v + \nu \cdot \mo)}  \tau(16 \mo/v)) \phi
\ene
and 
\beq \label{cansa5}
\psi^{(1)} = \chi_{B(\nu t; |t|/2 )} \psi,  \hspace{1cm} \psi^{(2)} =  \chi_{B(\nu t; |t|/2 )^c} \psi.
\ene
We have that 
\beq \label{cansa6}
\| \psi^{(1)} \| \leq C \|  \chi_{B(\nu t; |t|/2 )} h(x)\|\cdot \Big \|  \frac{1}{ \big ((\mo + v\nu)^2 + m^2 \big )^{1/2} } \langle \mo \rangle^{- 1} \Big \| \cdot \| \phi \|_{{\bf H}^{1}(\mathbb{R}^3)} \leq C
\frac{1}{v(1 + |t|)^\zeta}  \| \phi \|_{{\bf H}^{1}(\mathbb{R}^3)}.   
\ene
Now we estimate $\psi^{(2)}$ using Remark \ref{rpento}:  
\begin{align} \label{cansa7}
\| \psi^{(2)} \| \leq & \Big [\| \chi_{B(\nu t; |t|/2 )^{\displaystyle{c}}}\: \cdot \: e^{- i t ( v + \nu \cdot \mo)}  \tau(16 \mo/v)) \chi_{B(0; |t|/8)}  \langle x \rangle^{- 2 l }\| + \| \chi_{B(0; |t|/8)^{\displaystyle{c}}}  \langle x \rangle^{-2 l }\| \Big]
\\ \notag & \cdot \Big \| \langle x \rangle^{2 l } \frac{1}{ \big ((\mo + v\nu)^2 + m^2 \big )^{1/2} } \phi   \Big \|
\\ \notag \leq & C \frac{1}{v(1 + |t|)^{2l }}  \| \phi \|_{{\bf H}_{\langle x \rangle^{4 l}}^1(\mathbb{R}^3)}, 
\end{align}
where we use the procedure in \eqref{he6}-\eqref{noquier}. Eq. \eqref{cansa1} follows from \eqref{cansa4}-\eqref{cansa7}. 
\bull

\newpage

\begin{figure}\label{fig1}
\begin{center}
\includegraphics[width=17cm]{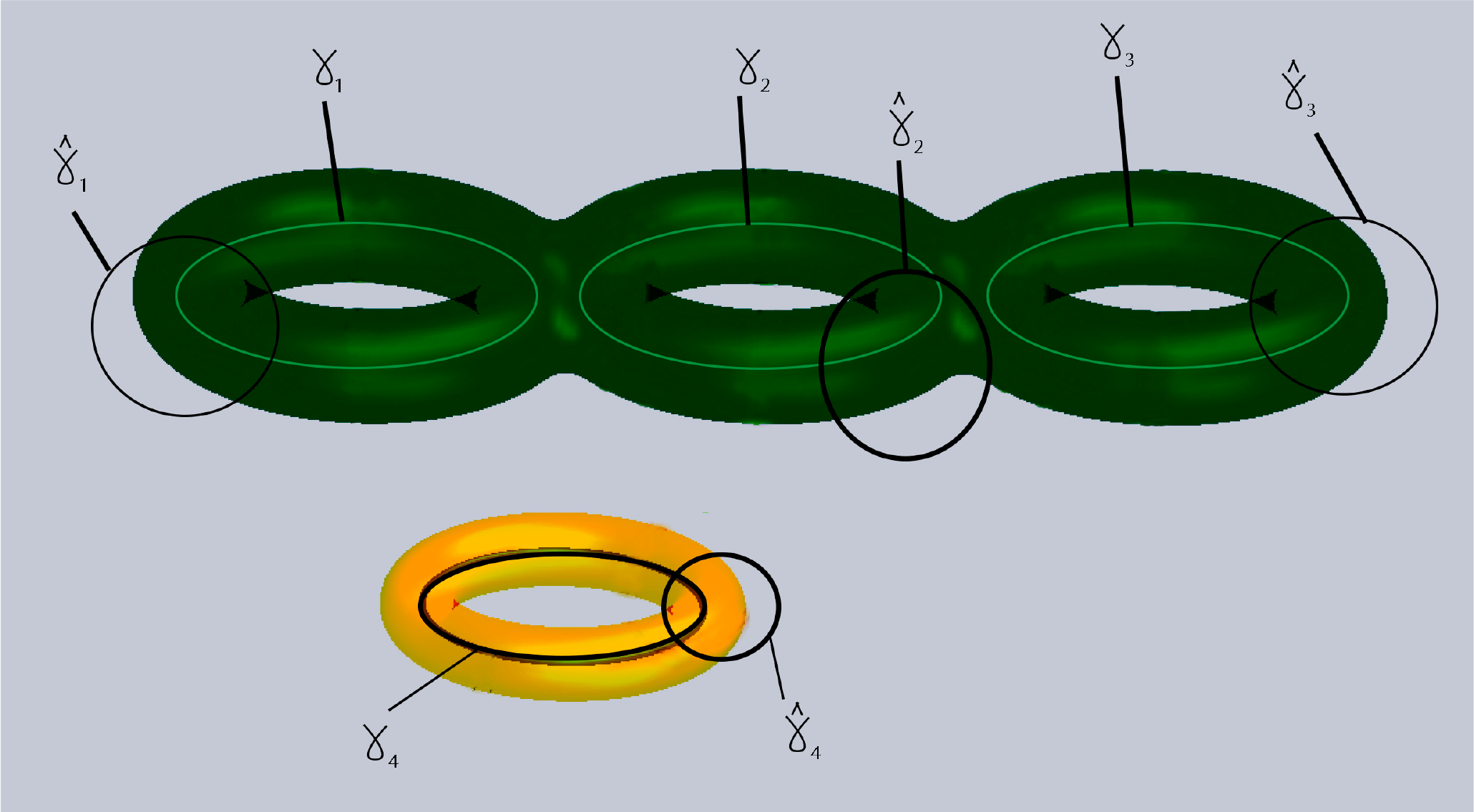}
\caption{The magnet  $ K= \cup_{j=1}^L  K_j \subset \ere^3$  where
$K_j$ are handlebodies, for
every $j \in \{ 1, \cdots, L \}$. The exterior domain,  $\Lambda:= \ere^3
\setminus K$. The curves $\gamma_k, k=1,2,\cdots m,$ are a basis of
the first singular homology group of $K$ and the curves
 $\hat{\gamma}_k, k=1,2,\cdots m,$ are a basis of the first singular homology group of $\Lambda$. }
\end{center}
\end{figure}
\newpage
\begin{figure}\label{fig2}
\begin{center}
\includegraphics[height=13cm,width=15cm]{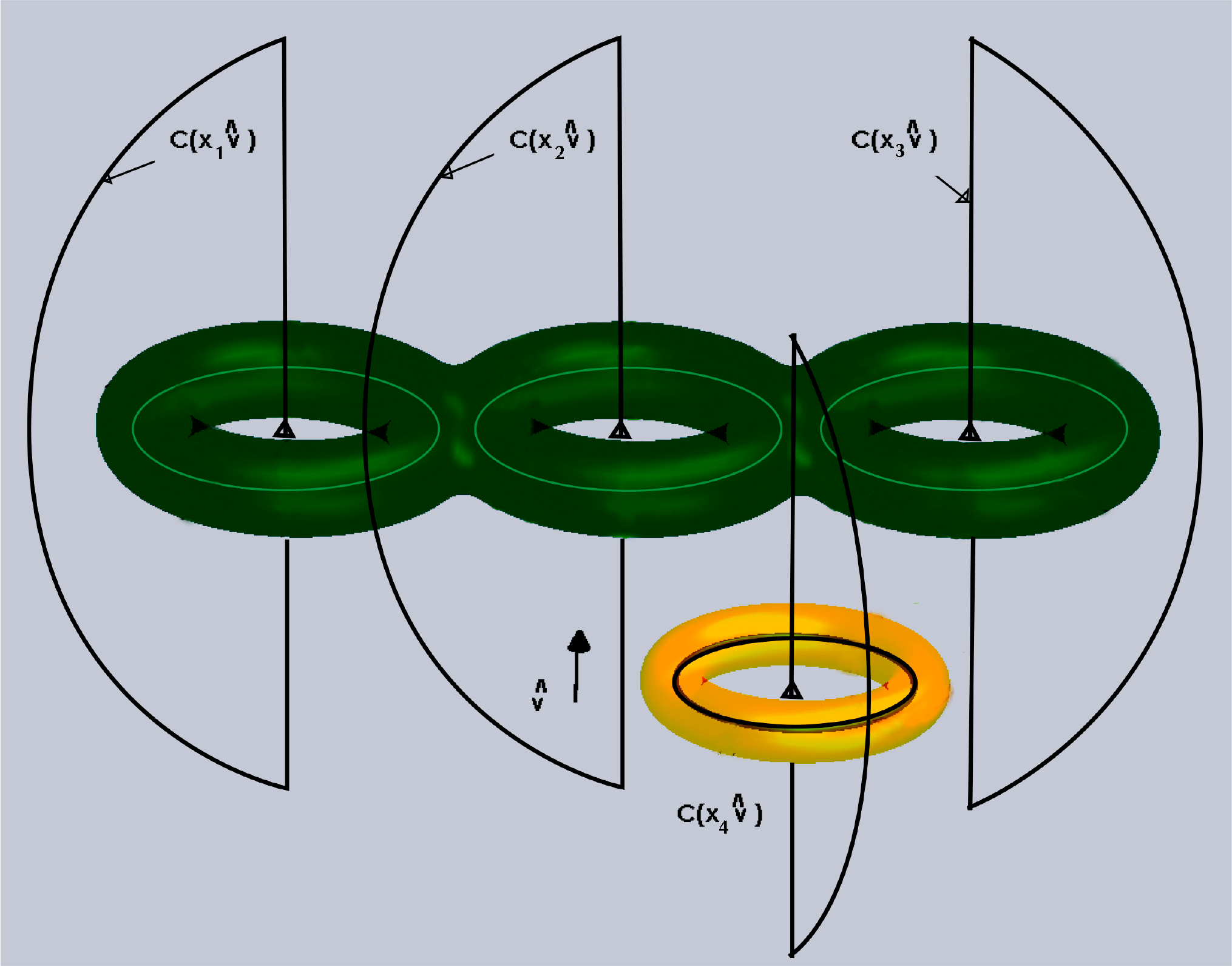}
\caption{The curves $c(x,\hv)$.}
\end{center}
\end{figure}

\end{document}